\documentclass{aa}

            \providecommand{\arcsec}{\ensuremath{^{\prime\prime}}}
            \providecommand{\arcmin}{\ensuremath{^{\prime}}}
        
            \usepackage{graphicx}  
            
            \usepackage{subfig} 
                
            \usepackage{multirow}
                
            \usepackage{txfonts} 

            \usepackage{natbib}
            \bibpunct{(}{)}{;}{a}{}{,} 
            
            \newcommand{\furl}[1]{\footnote{\url{#1}}}
    
            \newcommand{\FigDir}{./}
            \newcommand{\FigRootA}{\FigDir XPF_1D_JoinedModel_paper_plots}

    \usepackage{xspace}
    
    \newcommand{\xmm}{\textsl{XMM-Newton}}
    \newcommand{\planck}{\textsl{Planck}}
    \newcommand{\chandra}{\textsl{Chandra}}
    
    \newcommand{\xbootes}{\textsl{XBOOTES}}
    \newcommand{\xxl}{\textsl{XXL}}

    \newcommand{\Sref}[1]{Sect.~\ref{#1}}
    \newcommand{\Aref}[1]{Appendix~\ref{#1}}
    \newcommand{\Fref}[1]{Figure~\ref{f:#1}}
    \newcommand{\Tref}[1]{Table~\ref{#1}}
    \newcommand{\Eref}[1]{Eq.~\ref{#1}}
    \newcommand{\Eqref}[1]{Eq.~\eqref{#1}}
    
    \newcommand{\NH}{\ensuremath{N_\mathrm{H}}}
    \newcommand{\Mpch}{\ensuremath{\mathrm{Mpc}\,h^{-1}}}
    \newcommand{\Mpc}{\ensuremath{\mathrm{Mpc}}}
    \newcommand{\kpc}{\ensuremath{\mathrm{kpc}}}
    
    \providecommand{\sun}{\ensuremath{\odot}}       
    \newcommand{\msun}{\ensuremath{\mathrm{M_{\sun}}}}

    \newcommand{\APEC}{\texttt{APEC}}

    \newcommand{\LogNLogS}{\ensuremath{\log N - \log S}}
    
    \newcommand{\map}[1]{\ensuremath{\textit{\textbf{#1}}}}
    \newcommand{\outFOV}{\ensuremath{\mathrm{outFOV}}}
    \newcommand{\inFOV}{\ensuremath{\mathrm{inFOV}}}
    
    \renewcommand{\d}{\ensuremath{\mathrm{d}}} 
    
    \newcommand{\eg}{e.g.,\xspace}
    \newcommand{\ie}{i.e.\xspace}   

    \newcommand{\citeg}[1]{\citep[\eg{}][]{#1}}

        \newcommand{\KolodzigIII}{Kolodzig et al. 2021b, in prep.,} 
        \newcommand{\KolIII}{KOL21} 
        
        \defcitealias{PlanckEarlyIX2011}{PLIX2011}
        \newcommand{\PlXI}{\citetalias{PlanckEarlyIX2011}}

\begin{document}

    \title{X-ray analysis of the Planck-detected triplet-cluster system PLCK G334.8-38}
    
    \author{
            Alexander~Kolodzig\inst{\ref{inst1}}\and
            Nabila~Aghanim\inst{\ref{inst1}}\and
            Marian~Douspis\inst{\ref{inst1}}\and
            Etienne~Pointecouteau\inst{\ref{inst2}}\and 
            Edouard~Lecoq\inst{\ref{inst1}}
    }       
    \authorrunning{A. Kolodzig et al. (2O21)} 
    \institute{
        Universit\'e Paris-Saclay, CNRS, Institut d'astrophysique spatiale (IAS), 91405, Orsay, France,
            \email{alex@kolodzig.eu}\label{inst1}
            \and
            IRAP, Universit\'e de Toulouse, CNRS, CNES, UPS, Toulouse, France \label{inst2}
    }       
    
    \date{Received xx.xx.2O21; accepted xx.x.2O21}
    
    \abstract{ 
    We conducted an X-ray analysis of one of {the two} \planck{}-detected triplet-cluster systems, \texttt{PLCK G334.8-38.0}, with a $\sim100$~ks deep \xmm\ data.
    We find that the system has a redshift of $z=0.37\pm{0.01}$
    but the precision of the X-ray spectroscopy for two members is too low 
    to rule out a projected triplet system, demanding optical spectroscopy for further investigation.
    In projection, the system looks almost like an equilateral triangle with an edge length of $\sim2.0\,\Mpc$, but masses are very unevenly distributed ($M_{500} \sim [2.5,0.7,0.3] \times 10^{14}\,\mathrm{M_{\odot}}$ from bright to faint).
    The brightest member appears to be a relaxed cool-core cluster and is more than twice as massive as both other members combined. 
    The second brightest member appears to be a disturbed non-cool-core cluster and the third member was too faint to make any classification.
    None of the clusters have an overlapping $R_{500}$ region and no signs of cluster interaction were found; however, the \xmm\ data alone are probably not sensitive enough to detect such signs, and a joint analysis of X-ray and the thermal Sunyaev-Zeldovich effect (tSZ) is needed for further investigation, which may also reveal the presence of the {warm-hot intergalactic medium} (WHIM) within the system.
    The comparison with the other \planck{}-detected triplet-cluster-system shows that they have rather different configurations,
    suggesting rather different merger scenarios, under the assumption that they are both not simply projected triplet systems.
    } 
                        
            
    \maketitle


\section{Introduction} \label{s:intro}
In the standard paradigm, gravitation drives structure formation in a hierarchical process, and makes dark matter the ``scaffolding'' of the {cosmic web}. 
At first order, ordinary matter follows the dark matter distribution. 
This has long been demonstrated by theory and numerical simulations of structure formation. 
However, the role of baryons is much more complex in the process and the details of the physical processes governing this component remain to be understood.
In the local Universe, it is expected that the vast majority of baryons {($\gtrsim80\%$)} have been heated under the action of gravity to temperatures above $10^5\,\mathrm{K}$, and thus have not condensed into stars \citeg{Cen1999,Roncarelli2012,Kravtsov2012,Dolag2016,Martizzi2019}. 
This makes studying the ``hot'' Universe a crucial step towards understanding the formation and evolution of cosmic structures.

Important endpoints of structure formation processes are clusters of galaxies, reaching total masses above $10^{14}\,\msun$.
Most of their baryonic matter is in the form of a hot ($T \gtrsim 10^7\,\mathrm{K}$), tenuous plasma, the {intracluster medium} \citep[ICM; \eg{}][]{Sarazin1988}, representing $80 - 90\%$ of {the baryonic mass of a cluster of galaxies} \citeg{Giodini2009}.
However, {within the local Universe} the ICM only accounts for {a small fraction of baryons ($\sim4\%$)}; {the majority ($\sim50\%$)} are expected to reside in the cosmic web in the form of a {warm-hot intergalactic medium} (WHIM), a cooler and, on average, less dense plasma ($T \sim 10^5 - 10^7\,\mathrm{K}$, $n_e \lesssim 10^4\mathrm{cm^{-3}}$) than the ICM \citeg{Cen2006,Cautun2014,Martizzi2019}.
By understanding the physical state of the hot, diffuse gas and how it gets accreted, heated, and virialized onto clusters of galaxies in order to be transformed from a WHIM state to an ICM state, 
we can also better understand its role in structure formation.

Ideal places to study such structure formation processes are multi-cluster systems, also called {superclusters} ({SCs}).
{These are the most massive structures of the cosmic web 
and represent the largest agglomeration of galaxies (of the order of $10 - 100\,\Mpch$)},
containing thousands of them and a few to dozens of groups and clusters of galaxies \citeg{Einasto1980,Oort1983}.
SCs are already decoupled from the Hubble flow but are not yet virialized and most of them will eventually collapse under the effect of gravity.
This makes present-day SCs the largest bound but not yet fully evolved  objects in the local Universe, 
and therefore they can also be seen as ``island universes'' \citeg{ArayaMelo2009}.

Our current best knowledge of hot gas from SCs comes from targeted observations of nearby, X-ray-bright {multiple-cluster systems}.
These range from simple merging cluster pairs, such as the A399--A401 pair \citep{Ulmer1979} with its well-studied inter-cluster filament \citeg{Akamatsu2017,Bonjean2018,Govoni2019}, 
to very complex SCs, such as Shapley \citep{Shapley1930}, {the} richest known SC in terms of X-ray emitting clusters
\citeg{Raychaudhury1991,deFilippis2005}.
Studies of those SCs have shown that the  hot gas between cluster members of a SC, also known as intra-SC medium (ISCM), is accreted mainly along filaments and gets heated to ICM-like temperatures via shocks and adiabatic compression at the cluster outskirts \citeg{Tozzi2000,Zhang2020a,Power2020}.
Hereby, the interaction and merging of clusters  serves as an important catalyst of these accretion and heating processes and gives rise to other heating processes caused by the creation of additional turbulent gas motions \citeg{Sarazin2002,Shi2020}.
This means that the most important gas physics are happening in the outskirts of  clusters, beyond their viral radius \citeg{Ryu2003,Molnar2009,Nelson2014,Simionescu2019b}.

{In this respect, it is important to not just capture the emission of the ICM but also of the ISCM when studying the hot gas of an SC.
The hot gas within SCs is visible in the X-rays (through thermal bremsstrahlung and line emission),
and at millimeter wavelengths through the thermal Sunyaev-Zeldovich effect (tSZ). 
However, as it is cooler and less dense, detecting and studying the ISCM is much more challenging than for the ICM \citeg{Werner2008}.
One way of tackling this challenge is via stacking analysis.
The stacking analysis by \citet{Tanimura2019b} using tSZ data from \planck\ for galaxy-detected SCs revealed that the ISCM accounts for a significant fraction of WHIM  ($\gtrsim10\%$), which further emphasises SCs as suitable structures with which to study the role of hot bayrons in structure formation.
Another promising avenue can be found in joint analyses with X-rays and tSZ of specific SCs,}
because X-rays are more sensitive to cluster cores, while the tSZ effect is more sensitive to cluster outskirts.
The combination of both observables allows us to constrain the distribution of several important physical parameters of the hot baryons over a large physical range  \citeg{Eckert2016a,Keruzore2020}.

Serendipitously, a cluster-candidate search with \planck{} data {revealed} two triplet-cluster systems with favorable system configurations for
{the joint analysis of a SC of this type.}
Both are located at redshifts where they are still sufficiently X-ray-bright and they are also sufficiently small such that we can observe the entire system within one field-of-view (FOV) of the X-ray observatory \xmm{}, which is very well suited to detecting emission from diffuse, hot gas.
The search was conducted with the data from the first  \planck{}  sky scan \citep{PlanckEarlyVIII2011} and was followed up with an \xmm\ observation campaign \citep[hereafter PLIX2011]{PlanckEarlyIX2011}. 
For one of the SCs (\texttt{PLCK G214.6+36.9}), located at $z \approx 0.45$, a joint X-ray--tSZ analysis study {was}
conducted \citep{Planck_Int_VI_2013}.
The other triplet-cluster system {discovered in \planck{} data}, \texttt{PLCK G334.8-38.0}, is the focus of the present paper.
In  \Fref{Maps_Masks}, we show its X-ray and tSZ emission and mark its three cluster members with the letters A, B, and C, which are ordered according to the X-ray surface brightness (hereafter XSB) from bright to faint.

The first \xmm\ observation ($\sim25$~ks, OBS-ID: \texttt{0656200701}, {DDT time}) of this {\planck{} cluster} candidate, hereafter referred to as the {shallow} observation,
led to its identification as a triplet-cluster system by \PlXI{}.
{A first} analysis was conducted to measure luminosity and spectroscopic temperature and redshift, which were used in conjunction with scaling relations from the literature to derive estimates of the size and mass of each cluster. 
These triggered an approximately four times deeper \xmm\ observation ({$\sim111$~ks,} OBS-ID: \texttt{0674370101}, PI: E. Pointecouteau), hereafter referred to as  the {deep} observation.

We want to use this \xmm\ data in conjunction with \planck{} and optical data to conduct a comprehensive multi-wavelength analysis
in order to constrain the cluster members of the system and their dynamics, and the presence and the properties of WHIM around them.
This work represents the first step in this comprehensive analysis.
        
    The paper is organized as follows:
    \Sref{s:DataProc} describes our procedure of processing \xmm\ data,
    \Sref{s:Ana} explains our data analysis, 
        which includes the surface brightness measurement (\Sref{s:1D_XSB_measure_tot}),
        the spectroscopic measurement (\Sref{s:SpecFit}), 
        the model description (\Sref{s:mod}),  
        the fitting procedure (\Sref{s:LL}),
        and the fit results (\Sref{s:results}),
        and in \Sref{s:Dis} we discuss them and make comparisons with the literature.
    We summarize our {findings} in \Sref{s:sum}.
    In the Appendix, we 
        provide further details of the 
            spectroscopy (\Aref{a:CXB_spec}), 
            profile projection onto the sky (\Aref{a:proj}),
            and fit results (\Aref{a:1DXfit}).

    {Throughout the paper,} we assume a flat $\Lambda$CDM cosmology with the following parameters \citep[based on][]{Planck_CosPara}:
    $H_0 = 67.7\,\mathrm{km\,s^{-1}\,Mpc^{-1}}$ ($h=0.70$),
    $\Omega_\mathrm{m} = 0.307$ ($\Omega_\Lambda = 0.691$),
    $\Omega_\mathrm{b} = 0.0486$.

        \begin{figure*}
        \centering        
        \subfloat{      
                \includegraphics[width=0.33\textwidth]{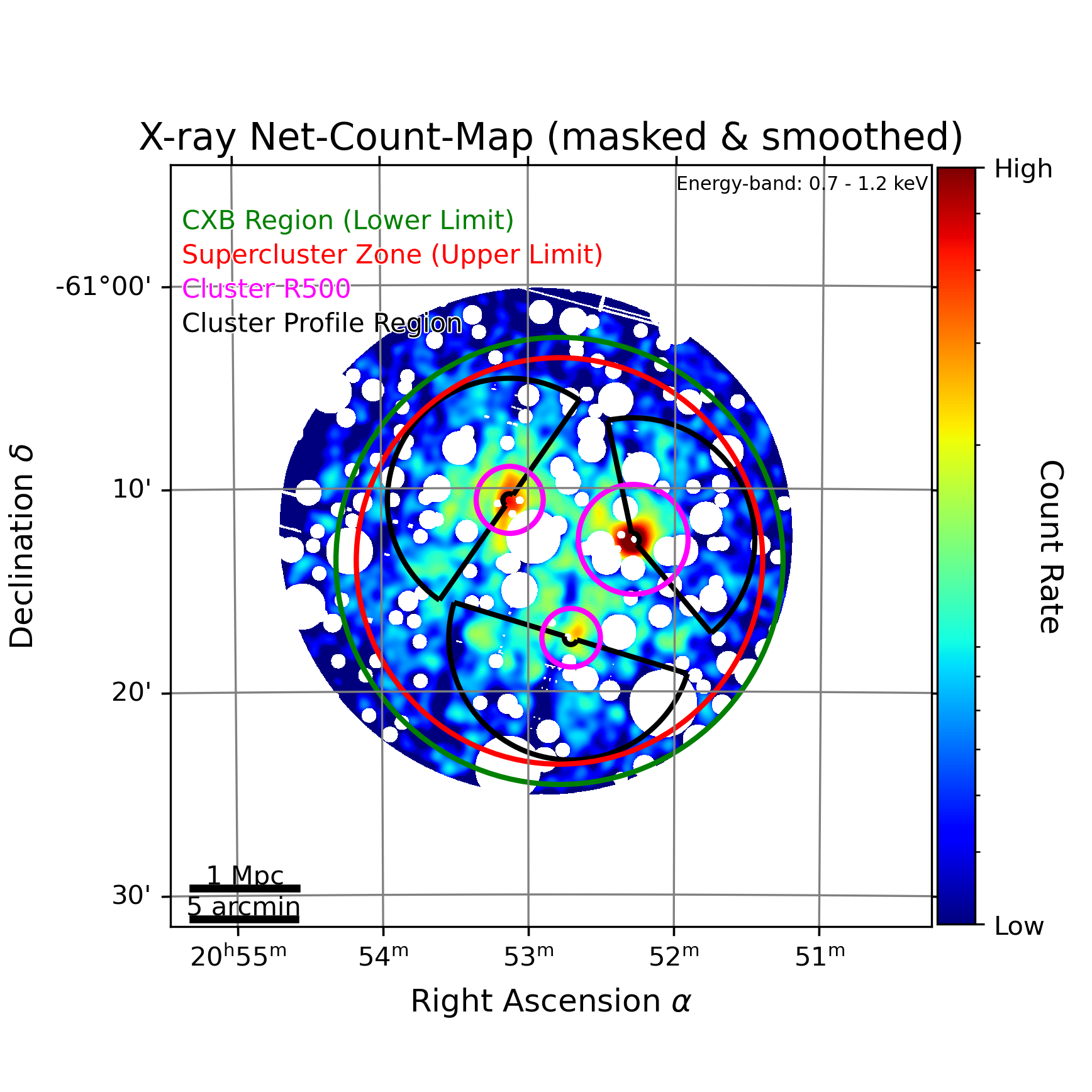} 
        }
        \subfloat{
                \includegraphics[width=0.33\textwidth]{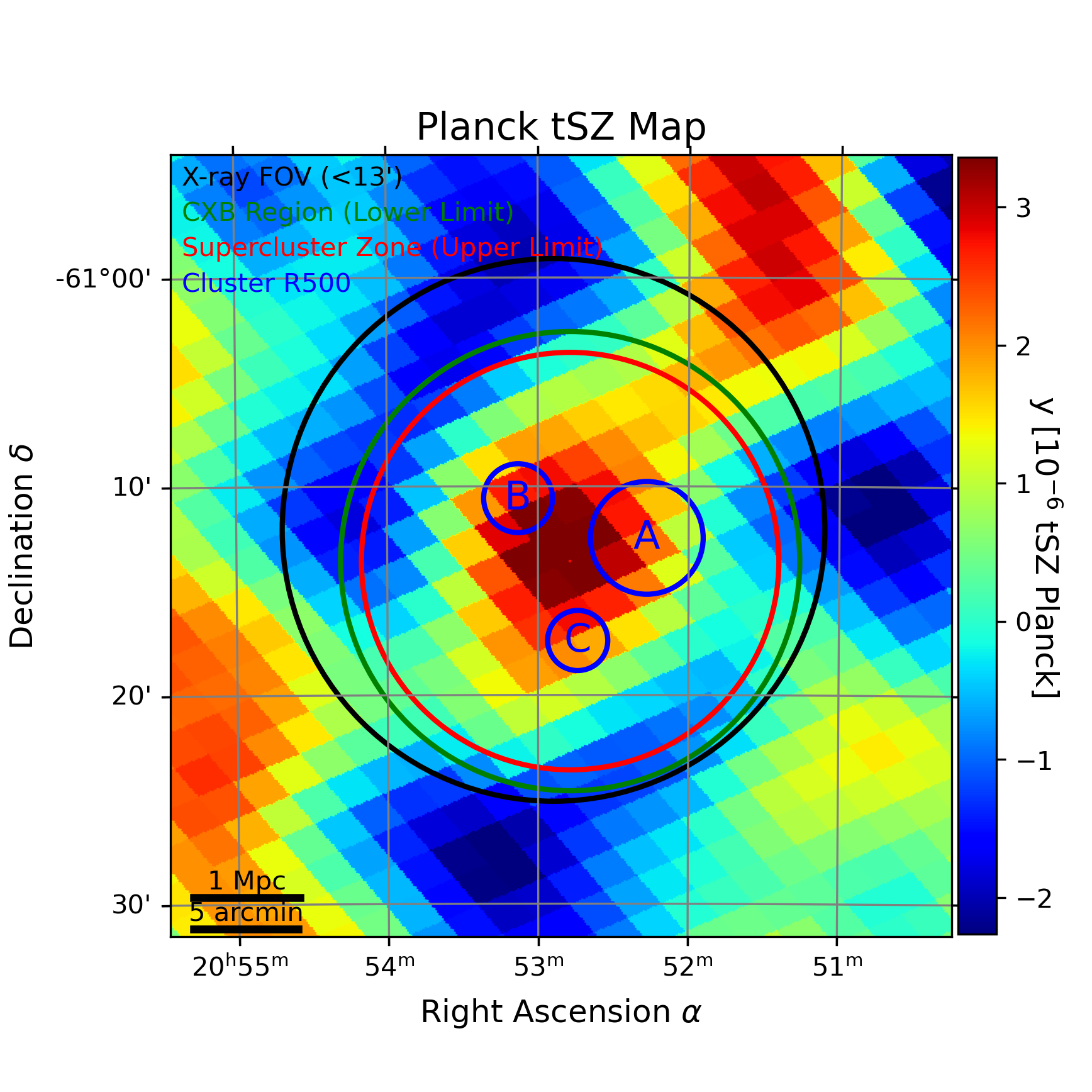} 
        }       
        \subfloat{
            \begin{minipage}[t][1\width]{0.33\textwidth}%
                    \includegraphics[width=\textwidth]{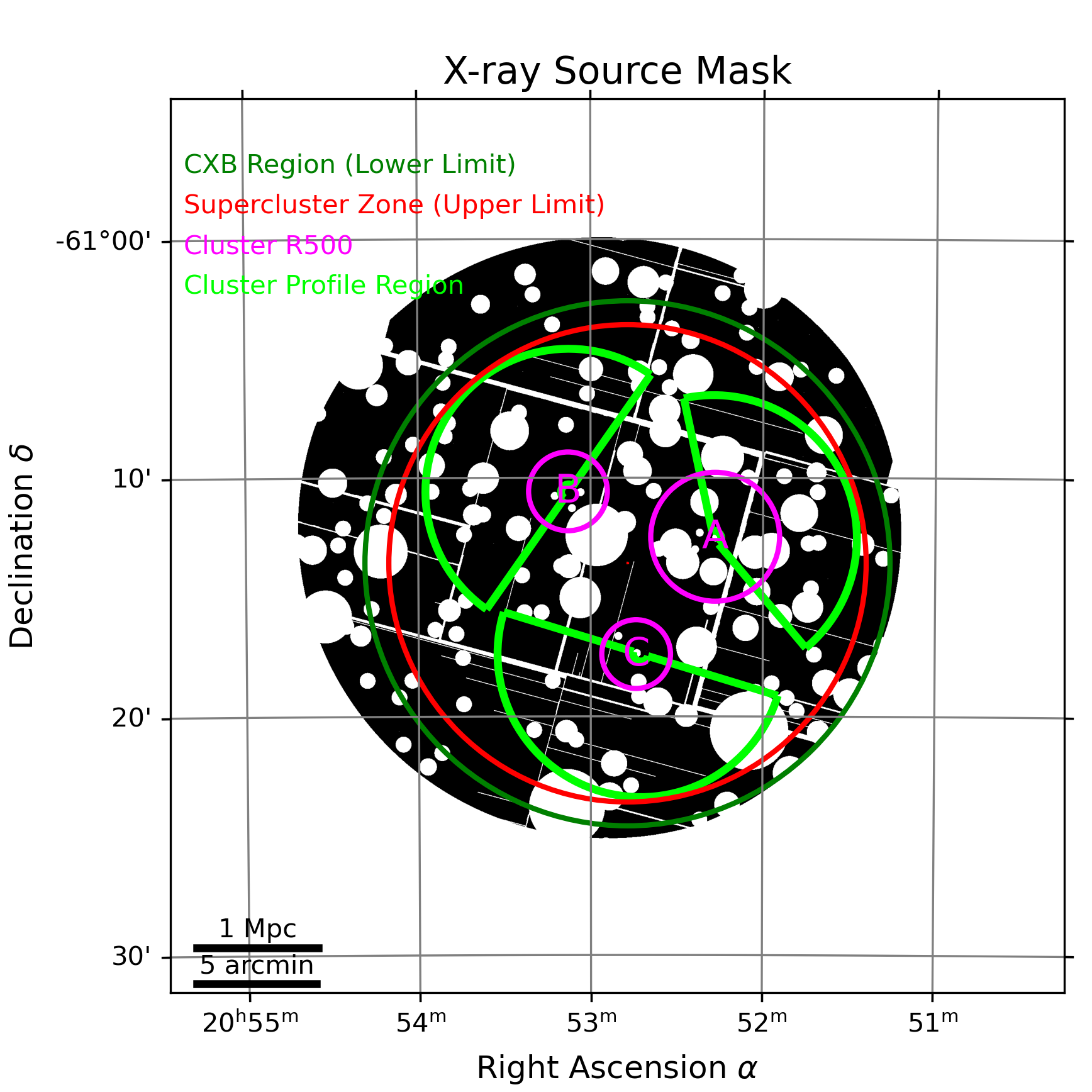}
        \end{minipage}
        }
        \caption{\label{f:Maps_Masks} 
                        X-ray and tSZ emission of the triplet-cluster system \texttt{PLCK G334.8-38.0}.
                        \emph{Left:} \xmm{} background-subtracted, detector-averaged count-rate map for $0.7-1.2\,\mathrm{keV}$ (masked and smoothed with a $15\arcsec$ wide Gaussian kernel).
                        \emph{Middle:} \planck{} tSZ signal \citep{Planck2015XXII}. 
                        \emph{Right:} X-ray source mask.
                        {Radial profiles are extracted from the cluster profile region
                        and the supercluster zone was designed to enclose all three of those regions.
                        The CXB region is used to study the astrophysical background.
                        Further details of the different regions are given in \Sref{s:regs}.}
                }
        \end{figure*}

\section{X-ray data processing} \label{s:DataProc}
        \subsection{Event filtering}  \label{s:evt}
        Here, we analyze the deep observation (OBS-ID: \texttt{0674370101}).
        We used the standard event-pattern-filter: \texttt{PATTERN <= 12} for EMOS and \texttt{PATTERN <= 4} for EPN.
        We used the standard event-filter \texttt{FLAG == 0} for events  within the \inFOV\ area
        and the standard event-filter \texttt{(FLAG \& 0x766aa000) == 0} for EMOS and \texttt{\#XMMEA\_EP} for EPN for events within the \outFOV\ area.
        For EMOS, we used the SAS task \texttt{emtaglenoise} 
        to detect and filter out CCDs, which are in an {anomalous state} or show high {electron noise} \citep{Kuntz2008}.
        For EPN, we applied the SAS task \texttt{epspatialcti} 
        to correct {the} event list {from} spatial variations in the charge-transfer-inefficiency, which is particular important for extended sources.
        {The remaining exposure time is [108,90]~ks for [EMOS,EPN], respectively.}

        \subsection{Background component treatment}
        \subsubsection{Particle background}  \label{s:IBKG}
        {To} remove the intervals of the observation contaminated by
        flares, a process called {deflaring},
        we use the 
        procedure of \KolodzigIII\ (hereafter \KolIII{}).
        { This filters the light curve of an observation in three consecutive steps of histogram clipping, 
        where the first step is tuned for the most obvious flares,
        the second step is tuned for weaker and longer flares using larger time bins,
        and the last steps is tuned for remaining flares by soft protons.
        These particles have energies smaller than a few hundred $\,\mathrm{MeV}$
        and can be funneled towards the detectors by the X-ray mirrors\furl{https://heasarc.gsfc.nasa.gov/docs/xmm/uhb/epicextbkgd.html}.
        This means that they are only detected within the \inFOV\ area but not for the \outFOV\ area.
        Hence, remaining flares of soft protons are detected with the light curve of the count-rate ratio between the \inFOV\ and \outFOV\ area.
        We {denote} those particles the {soft-proton background} (SPB),
        which represent one of the two major components of the \xmm{} instrumental background (hereafter IBKG).
        The other is caused by very energetic particles (with energies larger than some $100\,\mathrm{MeV}$) and is called the {high-energy-particle induced background} (HEB).
        {As SPB is mainly {flare residuals}, typically much brighter than the HEB,}
        {the deflaring procedure leads to a significant SPB reduction.}}
                
        Because of the deflaring, $\sim40-50\%$ of the exposure time was {not used in the analysis} ($\sim$[54,38]ks for [EMOS,EPN]). 
        More importantly, the fractional contribution of the SPB {with} respect to the total IBKG, estimated with eq. D2 of \KolIII{}, becomes zero in the $10-[12.0,15.0]\,\mathrm{keV}$ range for [EMOS,EPN], respectively, which {clearly shows} that, after {deflaring,} the observation contains a {negligible} quiescent contribution from the SPB \citeg{Leccardi2008,Kuntz2008}.
        
        \subsubsection{Astrophysical background} \label{s:ABKG}
        {In our work, the cosmic X-ray background (CXB) is defined as the accumulative emission of Galactic and extragalactic astrophysical fore- and background sources after masking out resolved sources.
        } 
        We identified resolved sources within our observations with the help of the 3XMM-DR8 catalog%
            \furl{http://xmmssc.irap.omp.eu/catalogue/3XMM-DR8/3XMM_DR8.html}  \citep{Rosen2016}.
        Following \KolIII{}, we use a radius of $30\arcsec$ as our default radius for the circular exclusion region for point-like
        sources, 
        which leads on average to a $\sim2\,\%$ fraction of residual counts of
        resolved point sources with respect to the total counts of the CXB. 
        For extended sources, two times their size {reported in} the 3XMM-DR8 catalog was used as the radius {of the exclusion area}.
        For the brightest sources (extended or point like), the radius was further increased until no significant residual emission was detected in an adaptively smoothed count-rate image. 
        For point sources within $1\arcmin$ of the core of a cluster, the circular exclusion region was reduced to $10\arcsec$.
        {We note that this only affects clusters A and C, because the core region of cluster B does not have any detected point sources within the 3XMM-DR8 catalog.}
        The resulting source mask is shown in the right panel of \Fref{Maps_Masks}.     
        
        \subsection{Region definitions} \label{s:regs}
        For our analysis, we need to define several sky and detector {regions shown} in \Fref{Maps_Masks}.
        Our \inFOV\ area is defined as a $13\arcmin$ circle centered on the on-axis point of the instrument. 
        This definition avoids the most outer parts of the \xmm{} full FOV, where the effective area is the smallest, {and} where the IBKG is the highest {with} respect to {the} astrophysical emission,
        {which can make the detection of (contaminating) sources unreliable \citeg{Chiappetti2013}}.
        The \outFOV\ area is used for {estimating} the IBKG and we adopt the definition within the XMM-ESAS scripts%
                \footnote{\label{ESAS}\xmm\ Extended Source Analysis Software, \url{https://www.cosmos.esa.int/web/xmm-newton/xmm-esas}}
        \citep{Snowden2004} for it.
        
        The {cluster region} 
        is defined as a circle of $6\arcmin$ radius {($\approx1.9\,\Mpc$)} centered on the X-ray emission peak of a cluster, 
        which is defined as {the count rate(IBKG-subtracted)-weighted barycentre of the cluster.}
        The {cluster profile region} is a subset of the cluster region, from which the temperature and XSB profiles are extracted.
        It is designed as a sector, which {excludes contributions from} the other clusters, as shown in \Fref{Maps_Masks} ({light green sectors in the} right panel).
        The opening angles of these sectors represent a compromise between maximizing the area of each cluster and minimizing the contaminating emission from the other clusters.
        The sector of cluster A has a smaller opening angle than the other clusters ($150^\circ$ instead of $180^\circ$), because it is the {brightest.}
        
        The {supercluster zone} (SCZ; {red circle in \Fref{Maps_Masks}}) describes a circle with a $10\arcmin$ radius. 
        {Its center is defined as $\alpha_\mathrm{SCC} = \mathrm{20^h\,52^m\,47.31^s}$ and $\delta_\mathrm{SCC} = \mathrm{-61^d\,13^m\,36.30^s}$, which was chosen to ensure that the SCZ encloses all three cluster regions.}
        The {CXB region} is used to 
        study the astrophysical background
        and is defined as the {annulus between $11\arcmin$ and $13\arcmin$}
        {(dark green and black circles in \Fref{Maps_Masks}) using the on-axis point as its center}.
        
\begin{figure*}
\begin{center} 
\resizebox{\hsize}{!}{\includegraphics{\FigDir 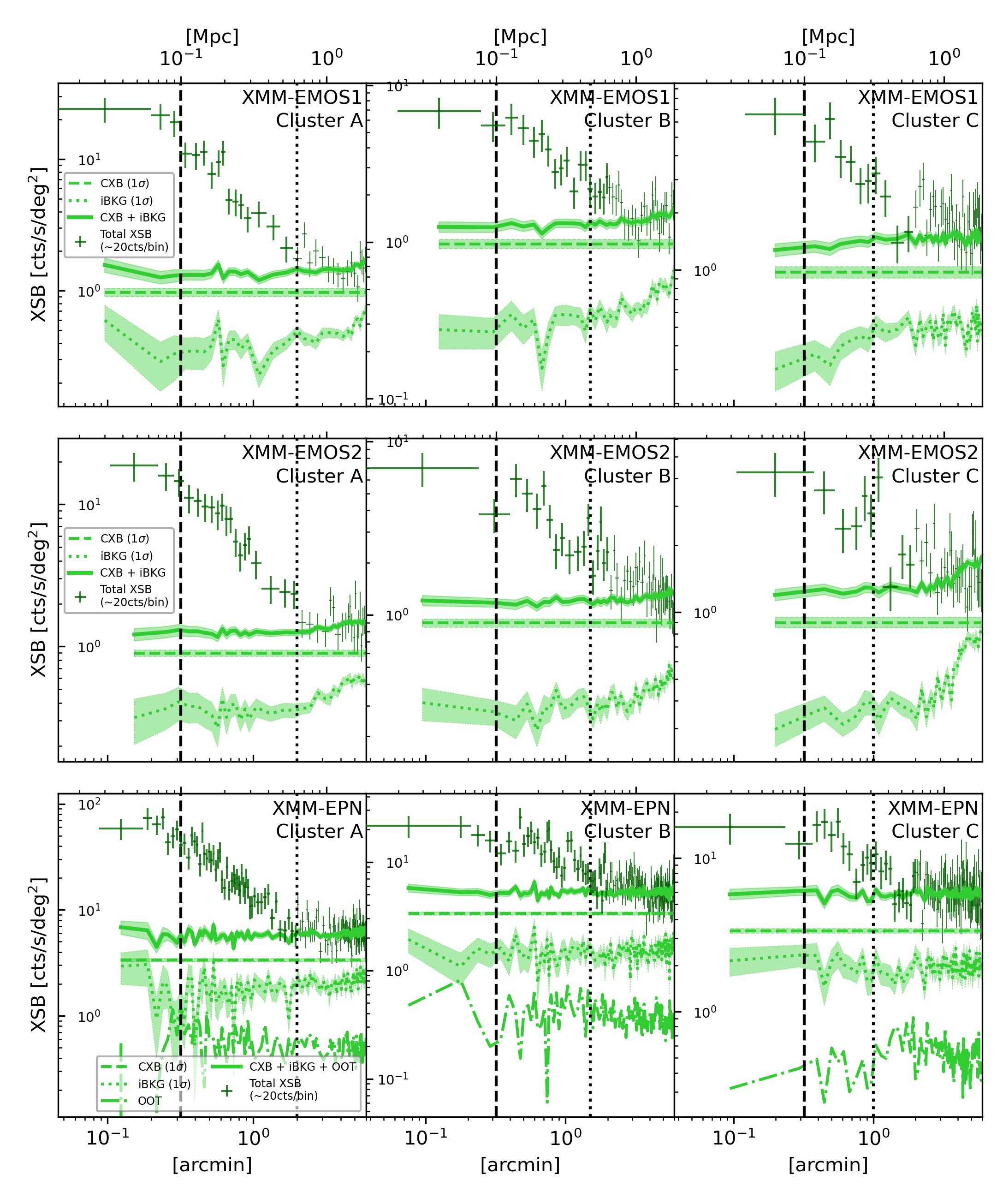}} 
\caption{\label{f:Total_XSB} 
        Measured total XSB profiles from the deep \xmm\ observation in the $0.7-1.2\,\mathrm{keV}$ band for each cluster and detector.
        Also shown are the independently measured background components, such as the astrophysical background (CXB, \Eref{eq:XSB_CXB}), the instrumental background (IBKG, \Eref{eq:IBKG}), and for EPN also the expected out-of-time events (OOT).
        For visualization purposes, the thickness of the green crosses is reduced for $>2\arcmin$.
        A logarithmic scale for the x-axis was used to highlight the cluster emission at small radii.
        {The dashed vertical line shows the upper limit of the core region ($<100\,\kpc$, \Sref{s:ExCore}).
        The scales between the dashed and dotted vertical lines are used for spectroscopy  (\Sref{s:SpecFit}).}
}
\end{center}            
\end{figure*}   
        
\section{Data analysis and modeling} \label{s:Ana}

    \subsection{Surface brightness measurement} \label{s:1D_XSB_measure_tot}
    \label{s:XSBprof} 
    {From the event file, we create a count map $\map{C}^\mathrm{T}$ [$\mathrm{cts}$]  for each {detector} with a resolution of $1.0\arcsec$ per pixel. 
    From the SAS task \texttt{eexpmap}, we computed the associate exposure $\map{E}$ [$\mathrm{s}$].}
        We note that $\map{E}$ corrects for various instrumental effects, such as mirror vignetting, spatial quantum efficiency, and filter transmission.
        The maps are created for the $0.7-1.2\,\mathrm{keV}$ band,
        because it represents the best {compromise} between \xmm{} high effective area and low background (CXB and IBKG), 
        when dealing with the extended emission of clusters \citeg{Eckert2016a}.

        The binning of the radial XSB profile of each cluster is adaptive in order to obtain $20\pm1$ total counts in each profile bin for a given detector, which ensures well-behaved statistics in each bin.    
        {The resulting XSB profiles are shown in \Fref{Total_XSB} for each cluster and detector.
        These illustrate that each XSB profile is a linear combination of the contribution from the cluster, the CXB (dashed lines), and the IBKG (dotted curves).
        For EPN, we also show the expected contribution from out-of-time events\furl{https://www.cosmos.esa.int/web/xmm-newton/sas-thread-epic-oot} (dot-dashed lines).}

                \subsubsection{Instrumental background} \label{s:XSBprof_IBKG}
                {For each detector, the XSB of the IBKG for the profile bin $b$ is computed as follows:
                        \begin{align} 
                                S^\mathrm{IBKG}_b & = \alpha^\mathrm{Clu} \; \dfrac{C^\mathrm{MB}_b}{\Omega_b \; \langle t_b \rangle }  \text{ ,} \label{eq:IBKG} \\
                                \sigma\left(S^\mathrm{IBKG}_b\right) & = \alpha^\mathrm{Clu} \;
                                \dfrac{\sqrt{C^\mathrm{MB}_b}}{\Omega_b \; \langle t_b \rangle } 
                                \notag
                                \text{ ,}
                        \end{align}
                where $\Omega_b$ and $\langle t_b \rangle$ are the surface area and average exposure time of the profile bin $b$.
                Here, we make use of the master-background-count map $\map{C}^\mathrm{MB}$,
                which was constructed out of the \emph{Filter-Wheel-Closed} observations of \xmm{}\furl{https://www.cosmos.esa.int/web/xmm-newton/filter-closed}
                following the description of \KolIII\ (appendix C).}
                
                The scaling factor $\alpha^\mathrm{Clu}$ in \Eqref{eq:IBKG} is computed as:
                        \begin{align} 
                                \alpha^\mathrm{Clu} & = 0.5 \left( \alpha^\outFOV + \alpha^\mathrm{Clu}_\mathrm{PB}\right) \text{ ,} \label{eq:scale}
                        \end{align}
                {where $\alpha^\outFOV$ scales $\map{C}^\mathrm{MB}$ to the total-count map $\map{C}^\mathrm{T}$ using the counts within the \outFOV\ area for the same energy band as the profile
                and $\alpha^\mathrm{Clu}_\mathrm{PB}$ scales $\map{C}^\mathrm{MB}$ to $\map{C}^\mathrm{T}$ using the counts within the particle band%
                        \footnote{The particle band is the energy band for a given detector, where only photons of the IBKG can be detected. 
                        For [EMOS,EPN], we use $10-[12.0,15.0]\,\mathrm{keV}$, respectively.}
                of the same detector area as the entire profile of the corresponding cluster.}
                Both methods are described in more detail in \KolIII\ (Appendix B).             
                
                \subsubsection{Astrophysical background} \label{s:XSBprof_CXB}
                {The CXB contribution is estimated from the CXB region (defined in \Sref{s:regs}) of the total-count map $\map{C}^\mathrm{T}$ as follows: 
                        \begin{align} 
                                S^\mathrm{CXB} & = \dfrac{1}{\Omega^\mathrm{CXB}} \sum_p^N  \dfrac{ M^\mathrm{CXB} }{ E } 
                                        \left( C^\mathrm{T} - \alpha^\mathrm{CXB} \, C^\mathrm{MB}  \right) \text{ ,} \label{eq:XSB_CXB} \\
                                \sigma\left(S^\mathrm{CXB}\right) & =  \dfrac{1}{\Omega^\mathrm{CXB}} 
                                        \left[ \sum_p^N  \dfrac{ M^\mathrm{CXB} }{ E^2 } 
                                        \left( C^\mathrm{T} + \left( \alpha^\mathrm{CXB}\right)^2 \, C^\mathrm{MB}   \right)
                                        \right]^{\frac{1}{2}} \notag \text{ ,} 
                        \end{align}             
                where $\map{M}^\mathrm{CXB}$, $\Omega^\mathrm{CXB}$, and $\alpha^\mathrm{CXB}$ are the mask, the surface area, and the $\map{C}^\mathrm{MB}$-scaling factor of the CXB region.
                Given this definition, it follows that $S^\mathrm{CXB}$ has the same value for all clusters for a given detector.
                In \Fref{Total_XSB}, we can see that the CXB (dashed lines) is the dominant background component 
                and for very large radii it dominates the entire emission.}
        
\begin{table*}
\begin{center}
\caption{\label{t:SpecFit_general} 
        Best-fit values and their $1\sigma$ levels of temperature $T$, redshift $z,$ and metallicity $m_Z$.
        }
\begin{tabular}{l|ccc|ccc|ccc}
\hline
 &  \multicolumn{3}{c}{Cluster A} &  \multicolumn{3}{c}{Cluster B} &  \multicolumn{3}{c}{Cluster C} \\
  & i                         & ii                        & iii           & i                      & ii                  & iii                 & i                      & ii                  & iii                    \\
\hline
 $T$ [kev]   & $3.9\pm{0.3}$             & $3.8\pm{0.3}$             & $3.9\pm{0.3}$ & $1.8_{-0.2}^{+0.3}$    & $2.0_{-0.1}^{+0.2}$ & $2.2_{-0.1}^{+0.2}$ & $1.4\pm{0.2}$          & $1.3_{-0.1}^{+0.2}$ & $1.4\pm{0.1}$ \\
 $z$   & $0.37\pm{0.01}$ & $0.37\pm{0.01}$ & $0.365$       & $0.26\pm{0.03}$        & $0.27\pm{0.02}$     & $0.365$             & $0.33_{-0.05}^{+0.04}$ & $0.31\pm{0.05}$     & $0.365$                \\
 $m_Z$ & $0.3\pm{0.1}$    & $0.3$                     & $0.3$         & $0.2\pm{0.1}$ & $0.3$               & $0.3$               & $0.4\pm{0.2}$          & $0.3$               & $0.3$                  \\
\hline
\end{tabular}
\tablefoot{
    Values obtained by modeling the X-ray energy spectrum of a broad profile region of each cluster (\Sref{s:Clu_Ave}).
    For the first fit (i), all three parameters were free, 
        for the second (ii) the metallicity was fixed to its default value,
        and for the third (iii) the metallicity and redshift were fixed to our default values (see \Sref{s:M_spec} and \ref{s:z_spec}).
        }
\end{center}
\end{table*}

\begin{figure*}
\begin{center} 
        \resizebox{\hsize}{!}{\includegraphics{\FigDir 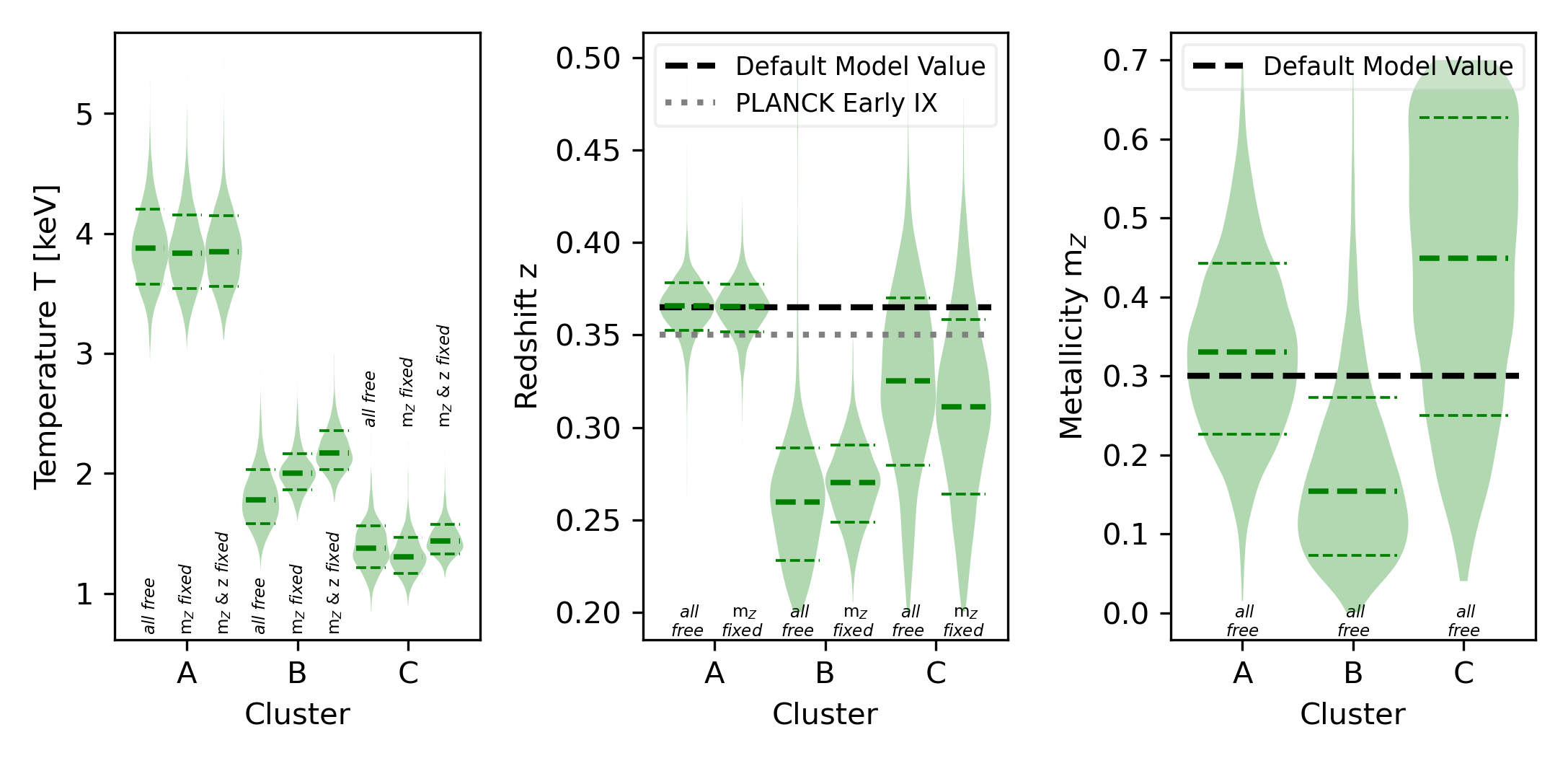}} 
            \caption{\label{f:SpecFit_general} 
            \label{f:SpecFit_AziBin} 
            \label{f:z_estimate} 
                Posterior distribution of temperature $T$, redshift $z$, and metallicity $m_Z$ after modeling the X-ray energy spectrum for a broad profile region of each cluster (\Sref{s:Clu_Ave}).
                Thick and thin dashed green lines show median and $1\sigma$ levels.
                {For the temperature and redshift, 
                we show different cases, where either all parameters are free (\textit{all free}) or at least one other parameter is fixed to the default value of our model, which is shown as a black dashed line in the middle and right panels.}
        {The gray dotted line in the middle panel shows the redshift estimate by \PlXI\ for cluster A based on the shallow \xmm\ observation.}        
        }
\end{center}            
\end{figure*}           
        
        \subsection{Spectroscopy} \label{s:SpecFit} 
        \label{a:SpecFit} 
        The spectrum extracted from a cluster is described with a one-temperature \APEC{}%
                \footnote{Using atomic database ATOMDB v3.0.9, \url{http://www.atomdb.org}.}
        model \citep{APEC}, 
        where its free parameters are the temperature, redshift, metallicity, and normalization.
        Galactic absorption is taken into account via the photoelectric absorption model \texttt{phabs}, for which we use 
        a hydrogen column density of $\NH=4.59\times10^{20}\mathrm{cm^{-2}}$ as determined by \citet{Kalberla2005}\furl{https://www.astro.uni-bonn.de/hisurvey/AllSky_profiles} for our \xmm\ observations
        and the solar metallicity of \citet{Anders1989}.
    For the spectral fitting we use the package 
        \textsc{XSPEC}%
            \furl{https://heasarc.gsfc.nasa.gov/docs/xanadu/xspec}
        \citep[v12.10.1f,][]{XSPEC} 
    in conjunction with the bayesian parameter estimation package 
        \textsc{BXA}%
            \furl{https://johannesbuchner.github.io/BXA}
        \citep[v3.3,][]{BXA}, 
        which requires the use of \texttt{cstat} for the fit statistic of 
            \textsc{XSPEC}
        and makes use of the nested sampling algorithm 
            \textsc{MultiNest}%
                \furl{https://github.com/farhanferoz/MultiNest}
        \citep[v3.10,][]{MultiNest2008,MultiNest2009,MultiNest2019}
        via the library 
            \textsc{PyMultiNest}%
                \furl{https://johannesbuchner.github.io/PyMultiNest/pymultinest.html}
            (v2.9).
        
        The CXB component of the energy spectrum is described with an \texttt{APEC + phabs(APEC + powerlaw)} model,
        which captures the emission of the local hot bubble, the Galactic halo, and extragalactic sources.
        For the \APEC\ models, the metallicity and redshift are set to unity and zero, respectively \citeg{Snowden2000,Henley2013},
        and the photon index of the power law is fixed to $1.46$ \citeg{Lumb2002,Moretti2003,DeLuca2004}.       
        The galactic absorption is the same as for the cluster model.
        The best-fit values of the remaining five free model parameters are determined from a spectral analysis of the CXB region of the observation and are used as fixed parameters, when fitting the energy spectrum of a cluster.
        They are listed in \Tref{t:CXBFit} and their posterior distributions are shown in \Fref{SpecFit_CXB}.
        The energy spectrum of the CXB region is shown in \Fref{CXB_spec_full}
        and the derived XSB of the three CXB model components are shown in \Fref{SpecFit_CXB_XSB}, which appear consistent with expectations.      
        
        Emission lines of the IBKG were identified and modeled {following} the XMM-ESAS documentation$^{\ref{ESAS}}$ but also considering the studies of
        \citet{Leccardi2008}, \citet{Kuntz2008}, \citet{Mernier2015}, and \citet{Gewering2017}.
        The continuum of the HEB is modeled with a broken power law.
        Best-fit values are obtained by fitting the HEB continuum
        to the energy spectrum of the master-background-count map (defined in \Sref{s:XSBprof_IBKG}) for the same detector area as the source  region; for example,\ a cluster profile bin or the CXB region.
        Thereby, energy ranges that contain known IBKG emission lines were ignored.
        The model parameters are fixed to these {best-fit} values, except for the normalization.
        For EPN, the HEB continuum model is used to describe the entire IBKG continuum, that is,\ including the SPB.
        For EMOS, the SPB continuum is described separately with a broken power law.
        We adopt the values of \citet{Leccardi2008} for the lower-energy slope and energy break parameter.
        The values of the high-energy slope and normalization are obtained from fitting the $5.0-11.2\,\mathrm{keV}$ band of the energy spectrum of the CXB region (\Fref{CXB_spec_full}), because this energy band has a negligible contribution from the CXB for our data.
        During this fit, the normalization of the HEB continuum is fixed via the XSB in the $10.0-11.2\,\mathrm{keV}$ band measured from the \outFOV\ area.
        When fitting the energy spectrum of a cluster profile bin or the CXB region,
        the normalization of the [HEB,IBKG] model for the [EMOS,EPN] detector remains the only free parameter of the IBKG continuum model.

                \subsubsection{Global values} \label{s:Clu_Ave}
                To estimate the average spectroscopic temperature, redshift, and metallicity
                for each cluster,
                we model the energy spectrum {over a broad profile region with respect to angular scale}.
                To maximize the precision of those spectroscopic estimates, we set the maximum radius of this region to the largest possible angular scale, 
                where the cluster XSB is still not significantly lower than the XSB of the CXB (estimated via \Eref{eq:XSB_CXB}).
                This leads to radii of $[2.0\arcmin,1.5\arcmin,1.0\arcmin]$ for clusters [A, B, C], respectively,
                which was revealed {later} in our analysis to {correspond} to $\sim[0.7,0.9,0.7] \times R_{500}$ (\Tref{t:DevPara}).
                For cluster A, the spectrum is shown in \Fref{ClA_spec_full}.
                {The limits are shown as dotted vertical lines in \Fref{Total_XSB}.}
                The cluster core ($<100\,\kpc$) was excluded from {this analysis}
                {to avoid including the emission of {potential cool cores}.}

                We show {our estimates for the redshift and metallicity parameters} together with the temperature estimates in \Tref{t:SpecFit_general} and \Fref{SpecFit_general}.
                {For the temperature and redshift, 
                we show different cases, where either all parameters are free (\textit{all free}) or at least one other parameter is fixed to its default value (black dashed line), which we define in the following sections.}
        
                \subsubsection{Redshift} \label{s:z_spec}
                We estimate the redshift of each cluster by modeling the X-ray energy spectrum of the broad profile region defined in  \Sref{s:Clu_Ave},
                where the metallicity was fixed to our default value ($m_Z=0.3$, see \Sref{s:M_spec}).
                The results of the modeling are shown {as \textit{$m_Z$ fixed} case} in \Fref{z_estimate} and \Tref{t:SpecFit_general}.
                For clusters A and C, we obtain redshift estimates of {$z=0.37\pm{0.01}$}
                and $z=0.33_{-0.05}^{+0.04}$, respectively.
                Both values are consistent with each other and also consistent with the value of $z=0.35$ estimated by \PlXI{} (gray dotted line in the middle panel of \Fref{z_estimate}).
                As cluster A is the brightest of the three clusters and its redshift estimate has the highest precision,
                we use {its best-fit value} ($z=0.365$)
                {in the following as the default value for our cluster emission models.} 

                For cluster B, we estimate a redshift of $z=0.27\pm0.02$, which stands in strong tension with the redshifts of the other clusters, especially cluster A {($\sim5\sigma$ tension)}.
                This value would suggest that cluster B is actually not a member of the triple-cluster system, 
                as it would be $\sim350\,\Mpc$ closer to the observer in the line-of-sight (LoS) direction than the other two members.
                However,
                our redshift estimate of cluster B is based on modeling the cluster energy spectrum,
                {which does not have any significant line emission from iron, with a one-temperature \APEC{} model.}
                {This makes the redshift estimate unreliable.}

                \subsubsection{Metallicity} \label{s:M_spec}
                {To estimate the metallicity,}
                we fixed the redshift to our default value ($z=0.365$, see \Sref{s:z_spec}).
                This revealed that with the given data, it is not possible to properly constrain the metallicity for clusters B and C.
                For cluster A, we obtain $m_Z=0.33_{-0.10}^{+0.11}$.
                It is worth noting that this estimate does not show a significant degeneracy with the temperature and normalization parameters. 
                This estimate is consistent with the common assumption of $m_Z=0.3$ for solar metallicity, which is based on previous cluster metallicity studies
                (\eg{} \citealt{Mernier2018}%
                 \footnote{We note that \citet{Mernier2018} used a different solar abundance \citep{Lodders2009}, which may introduce some bias.
                 Fortunately, the best-fit values of temperature and redshift do not change significantly when using their solar abundance instead,
                 \ie\ such a potential bias is much smaller than our measurement uncertainties.
                 Moreover, those best-fit values also do not change significantly when the metallicity becomes a free parameter, as shown in \Fref{SpecFit_general}.}).
                Hence, {in the following, the metallicity of our cluster emission models is fixed to $m_Z=0.3$.}
        
                \begin{figure*}
                \begin{center} 
                        \resizebox{\hsize}{!}{\includegraphics{\FigDir 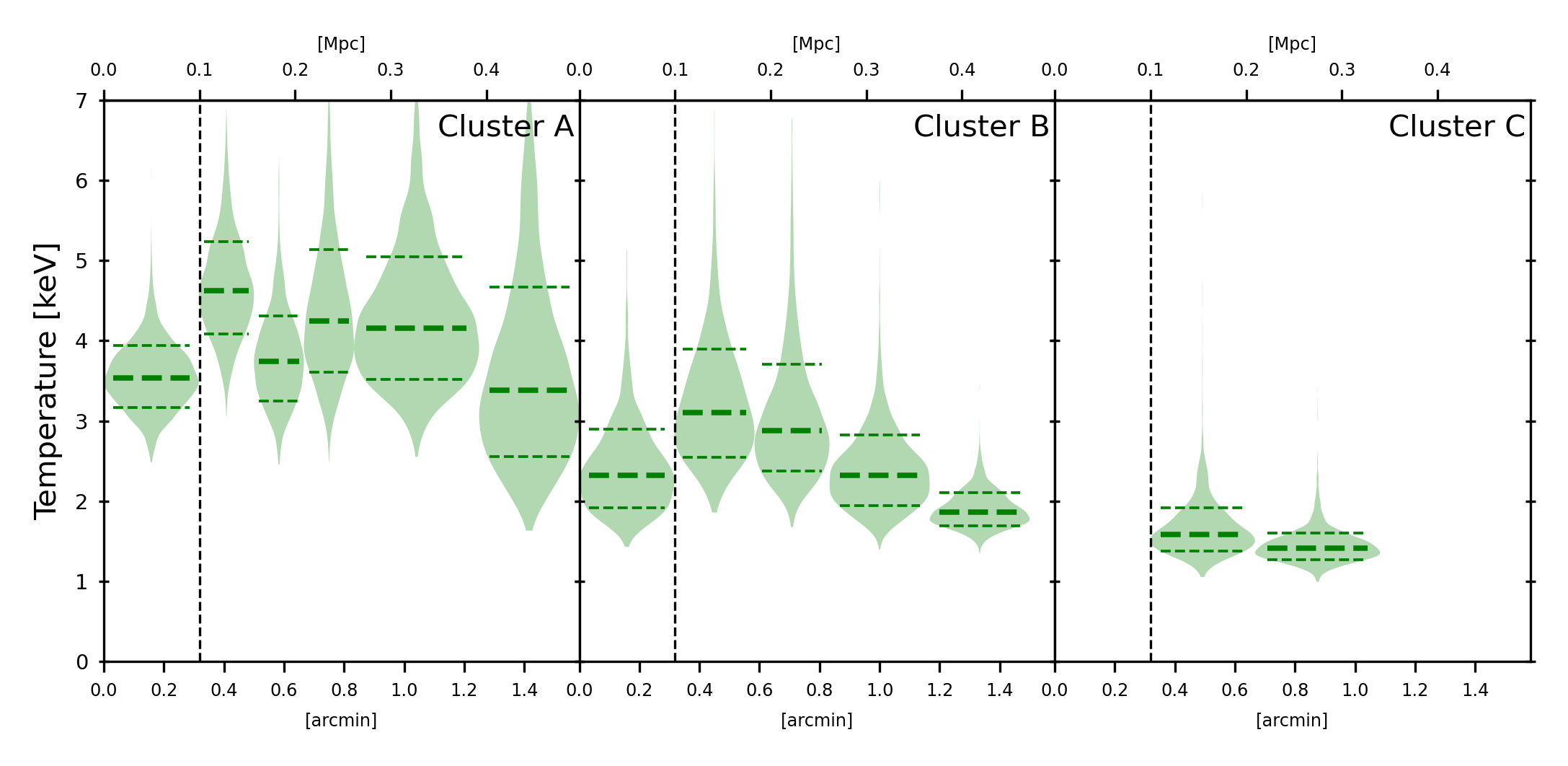}} 
                        \caption{\label{f:SpecFit_TempProf} 
                                Cluster temperature profiles:
                                Posterior distribution of the temperature after modeling the X-ray energy spectrum of consecutive profile bins of each cluster.
                                Thick and thin dash lines are median and $1\sigma$ levels.
                                The dashed vertical line shows the upper limit of the core region ($<100\,\kpc$, \Sref{s:ExCore}).
                        }
                \end{center}            
                \end{figure*}           
                \subsubsection{Measuring the cluster temperature profile} \label{s:Tprof}  
                In \Fref{SpecFit_TempProf}, we show our spectroscopic temperature profiles, 
                where redshift and metallicity were fixed to their default values ($z=0.365$ and  $m_Z=0.3$, see \Sref{s:z_spec} and \ref{s:M_spec}).
                The profiles are used for the joint fit in \Sref{s:LL}, where they are simultaneously fitted with the total XSB profiles. 
                We limit the profile bins to angular scales of [$95\arcsec$,$90\arcsec$,$65\arcsec$] for clusters [A,B,C], respectively,
                which was revealed {later} in our analysis to correspond to $\sim[0.5,0.9,0.8] \times R_{500}$ (\Tref{t:DevPara})
                since beyond this scale the amplitude of the cluster emission is at the same level as, or below, the amplitude of the CXB emission
                and our model of the CXB emission is not accurate enough to provide reliable measurements in this regime {for a single profile bin.}
        
        \subsection{Modeling} \label{s:mod}  
        \label{s:Mod_Prof}
        {To model the X-ray emission {of the gas} in and around clusters, we used {two different} parametric models directly projected onto the sky.}
        An important feature of this {forward-fitting approach} is that a model can be applied directly to the data.
        Both models provide a 3D radial profile for the electron number density $n_e$ and temperature $T$ of a cluster assuming spherical symmetry and a hydrostatic equilibrium with only three free parameters, {which will increase to five when including our model for the core region.}
        As both models use different underlying physical assumptions, they can be considered as independent.  
        
                \subsubsection{Isothermal beta-profile model}  \label{s:Mod_IsoBeta}
                {Our first model}
                assumes that the gas temperature is constant ($T(r_\mathrm{3D})=T_\mathrm{iso}$) and the density profile can be described with a beta-profile \citep{Cavaliere1976}: 
                \begin{align} \label{eq:BetaProf}
                        n_e(r_\mathrm{3D}) & = n_{e,0} \left[1+ \left( \frac{r_\mathrm{3D}}{r_c} \right)^2\right]^{- \frac{3}{2}\beta}  \text{ ,}
                \end{align}     
                where $r_\mathrm{3D}$ is the profile radius from the cluster center.
                Hence, 
                {the free model parameters are:}
                the temperature $T_\mathrm{iso}$, the core radius $r_c$, and the density profile normalization $n_{e,0}$.
                The slope $\beta$ {is kept} fixed to a value of two-thirds in our study. 
                
                This model neglects that the radial temperature of clusters declines beyond their central parts towards the outskirts \citeg{Markevitch1998,Ghirardini2019a}.
                The statistical quality of our data (see the uncertainties on our temperature measurements in \Fref{SpecFit_TempProf}) prevents us from a complex modeling of the temperature shape%
                    \footnote{A joint analysis of X-ray and tSZ may support the use of more complex models.}
                (\ie\ using an analytic function such as the one provided by \citealt{Vikhlinin2006}).
                We therefore restrict the complexity of this model via the assumption of an isothermal sphere, while for our second model, introduced in the following
                section, we use a different approach.
                Here, we purposefully exclude the cluster core, and we address it separately in \Sref{s:ExCore}.
        
                \subsubsection{Polytropic NFW-profile model}  \label{s:Mod_PolyNFW}
                Our second model assumes that the dark matter density profile can be described with a NFW profile \citep{Navarro1997}.
                Its free parameters are the characteristic radius $r_s$, where the profile slope changes, and the (dimensionless) concentration parameter $c_\Delta$.
                These define the reference radius $R_\Delta = c_\Delta \, r_s$,
                where the mean matter density of the cluster halo is $\Delta$ times the critical density of the Universe  $\rho_c$,
                resulting in an enclosed mass of $M_\Delta = 4/(3\pi) \, R_\Delta^3 \, \Delta \; \rho_c$.
                
                The model also assumes that the gas profile can be described with a polytropic gas model.
                Using the convention of \citet[hereafter KoSe01]{Komatsu2001},
                the electron number density profile is defined as: 
                        \begin{align}  \label{eq:PolyGasProf}
                                n_e(r_\mathrm{3D}) & = n_{e,0} \; \mathrm{y}_g(\tfrac{r_\mathrm{3D}}{r_s}) \text{ ,}
                        \end{align}
                and the gas temperature profile as: 
                        \begin{align}  \label{eq:PolyTProf}
                                T(r_\mathrm{3D}) & = T_0 \; \mathrm{y}^{\gamma-1}_g(\tfrac{r_\mathrm{3D}}{r_s}) \text{ ,}
                        \end{align}
                assuming the polytropic parameterization $P_g \propto n_e \, T \propto n_e^\gamma$ with the polytropic index $\gamma$.
                The dimensionless gas profile $\mathrm{y}_g$ is determined by solving the hydrostatic equilibrium equation (see eq.~19 of KoSe01)
                and {$T_0$ is computed via eq.~20 of KoSe01}.

                The parameterization of $n_e(r_\mathrm{3D})$ and $T(r_\mathrm{3D})$ adds two more model parameters: $n_{e,0}$ and $\gamma$.
                The polytropic index {is kept} correlated to $c_\Delta$ via the linear-fit formula of KoSe01 (eq.~25). 
                Hence, the three free model parameters are: $c_\Delta$, $r_s$, and $n_{e,0}$.
                Following the convention of KoSe01, we use $\Delta=200$.
                
        \subsubsection{Core region}   \label{s:ExCore}
        Each cluster may contain a cool core,
        which cannot be adequately described by either of our two simple models \citeg{Hudson2010,Komatsu2001}.
        To take such {a contribution} into account, 
        we split each radial profile model into a ``core'' region and a ``without-core'' profile (Out):
        \begin{align} 
                n_{e,0} & = 
                        \begin{cases}
                                n_{e,\mathrm{Core}} & \text{for } r_\mathrm{3D} < R_\mathrm{Core} \\
                                n_{e,\mathrm{Out}} & \text{for } r_\mathrm{3D} \geq R_\mathrm{Core}
                        \end{cases}
                    \label{eq:Ne_full} 
                    \;   \text{ ,} 
            \\
                T^\mathrm{(Full)}(r_\mathrm{3D}) & = 
                        \begin{cases}
                                T_\mathrm{Core} & \text{for } r_\mathrm{3D} < R_\mathrm{Core} \\
                                T(r_\mathrm{3D}) & \text{for } r_\mathrm{3D} \geq R_\mathrm{Core}
                        \end{cases}
                    \label{eq:T_3D_full}
                    \;   \text{ ,}
        \end{align}
        where
        the core-region upper limit is defined as $R_\mathrm{Core} = 100\,\kpc$  
        (which turns out to be $\sim[0.1,0.2,0.2] \times R_{500}$ for cluster [A,B,C], respectively)
        and $n_{e,\mathrm{Core}}$ and $T_\mathrm{Core}$ are two additional free parameters, 
        giving\ five free parameters in total for each cluster model.
        {Our model design can lead to a nonphysical discontinuity for the temperature and density between the core region and the without-core profile,
        especially for cool-core clusters.
        This is taken into account {when we interpret the results}.}
        We do not expect that this profile separation creates a strong bias in our understanding of the without-core profile, 
        because the core properties are typically not strongly correlated with it \citeg{Lau2015,McDonald2017}.
        
                \subsubsection{Radial profile models} \label{s:1D_XSB_model} 
                \label{s:1D_T_model} \label{s:1D_XSB_model_tot}
                Based on the density and temperature profile of each model,
                we compute its corresponding volume emissivity profile 
                as follows:
                \begin{align} \label{eq:VE_3D}
                        \epsilon_V(r_\mathrm{3D}) = & \,\, n_H(r_\mathrm{3D}) \, n_e(r_\mathrm{3D}) \, \notag \\
                                & \int\limits^{E_\mathrm{max}}_{E_\mathrm{min}} \d E \, \Lambda\left(T^\mathrm{(Full)}(r_\mathrm{3D}),m_Z,E,z\right)  \text{ ,}                      
                \end{align} 
                which follows the convention of the \APEC\ plasma code,
that is,\ using the hydrogen number density ($n_H$) instead of ion number density ($n_i$).
                We then compute the projected temperature profile in [keV] and the cluster XSB profile for each detector in instrumental units [cts s$^{-1}$ deg$^{-2}$] with the help of \xmm\ response files extracted from the on-axis point of the observation.
                The projected temperature profile $T_\mathrm{EW}$ is computed with the emission-weighted, projected 3D temperature profile $T^\mathrm{(Full)}(r_\mathrm{3D})$,
                and a cluster XSB profile $S^\mathrm{Clu}$ is computed with the projected volume emissivity profile $\epsilon_V(r_\mathrm{3D})$, using the Abel transform for the projection, as explained in \Aref{a:proj}, in both cases.
                
                {In order to compare the temperature profile model to the measurement,
                we have to fit it to the corresponding spectroscopic temperature profile (shown in \Fref{SpecFit_TempProf}).}
                We note that a comparison between emission-weighted and spectroscopic temperatures can be biased but only if the gradient of the underlying 3D temperature is strong \citeg{Vikhlinin2006b}.
                {Based on results from} \citet{Mazzotta2004}, we expect that such a bias is much smaller than our measurement uncertainties, 
                because our analysis reveals that each 3D temperature profile covers only a small temperature range within $R_{500}$ (\Sref{s:entropy}).
                
                Additional operations are necessary in order to compare the cluster XSB profile models with their corresponding measured XSB profiles (shown in \Fref{Total_XSB}). 
                First, the profile model is convolved with the \xmm{} PSF $f^\mathrm{Clu}_\mathrm{PSF}$ estimated for the detector position of the cluster center with the SAS task \texttt{psfgen}.
                Second, a CXB and a IBKG component are added, resulting in the following total XSB profile model of the profile bin $b$ for a given detector:
                \begin{align} \label{eq:1D_XSB_model}
                        S_b^\mathrm{T} & = \left(S^\mathrm{Clu} \ast f^\mathrm{Clu}_\mathrm{PSF}\right)(r_\perp^b) + S^\mathrm{CXB} + S^\mathrm{IBKG}_b   \text{ ,}
                \end{align}
                where $r_\perp$ is the profile radius from the cluster center within the projected plan, \ie\ perpendicular to the LoS.
                
                In \Eqref{eq:1D_XSB_model}, the IBKG component is defined with \Eqref{eq:IBKG}.
                The CXB component is modeled with a constant parameter $S^\mathrm{CXB}$ for the entire profile, which is a reasonable assumption.
                {We note that $S^\mathrm{CXB}$ can {include emission} from other cluster members and the ISCM.}
                To constrain it sufficiently well, it is important that XSB profiles
                {extend to radii} where the CXB emission dominates
                over {other emissions.}
                As  for our analysis 
                {this happens {beyond}}
                $\sim4\arcmin \approx 1.3\,\Mpc$ (see \Fref{Total_XSB}), 
                our profile region was defined to have a $6\arcmin$  radius ($\approx1.9\,\Mpc$) from the cluster center.
                The CXB component adds three more parameters to the XSB profile model (one per detector), 
                resulting in eight free parameters for the entire model.
                
                In order to use a Poisson likelihood during the fitting process, 
                the XSB profile model is converted from count rate [$\mathrm{cts\,s^{-1} deg^{-2}}$] to the sum of total counts [$\mathrm{cts}$] per profile bin as follows:
                \begin{align} \label{eq:1D_CTS_model}
                        \mathsf{C}^\mathrm{T}_b =  S_b^\mathrm{T} \; \Omega_b \; \langle t_b \rangle  + C_b^\mathrm{OOT} \text{ ,}
                \end{align}
                where $\Omega_b$ is the surface area,
                $\langle t_b \rangle$ the average exposure time,
                and $C_b^\mathrm{OOT}$ is the sum of counts expected from out-of-time events of EPN (and zero for the other detectors)
                for the profile bin $b$.
                We note that $C_b^\mathrm{OOT}$ is a negligible component for our analysis (see bottom panels of \Fref{Total_XSB}).

                \subsubsection{Fitting} \label{s:LL}
                For each cluster of the triple system, we fit one model simultaneously to the radial XSB profiles of all detectors (\Fref{Total_XSB})
                and the radial temperature profile (\Fref{SpecFit_TempProf}) via a maximum-likelihood estimation.
                The corresponding joint likelihood is computed as follows:
                \begin{align} 
                        \ln(\mathcal{L}) & = \sum_d^{N_\mathrm{XXM}} \ln\left(\mathcal{L}^\mathrm{(XSB)}_{d}\right) + 
                        \ln\left(\mathcal{L}^\mathrm{(T)}\right)  \text{ ,} \label{eq:JonidLL}
                \end{align}
                which sums up the logarithmic likelihoods of the XSB profiles and the temperature profile.
                {$N_\mathrm{XXM}=3$} is the number of \xmm\ detectors. 
        The logarithmic likelihood of the XSB profile of detector $d$ is computed as follows:
                \begin{align} 
                        \ln\left(\mathcal{L}^\mathrm{(XSB)}_{d}\right) & = \sum_b^\mathrm{(XSB)}  C^\mathrm{T}_b \; \ln\left( \mathsf{C}^\mathrm{T}_b  \right) - \mathsf{C}^\mathrm{T}_b \text{ .} \label{eq:C_LL} 
                \end{align}
                Here, we compute the logarithmic probability of measuring the sum of total counts $C^\mathrm{T}_b$ for the XSB profile bin $b$ of detector $d$
                given our model $\mathsf{C}^\mathrm{T}_b$ (\Eref{eq:1D_CTS_model}).
                The probability is computed with the Poisson distribution in natural logarithmic form, where all nonmodel-dependent terms are ignored because they are not relevant for the maximum-likelihood estimation. 
                The logarithmic likelihood of the temperature profile is computed as follows:
                \begin{align} 
                        \ln\left(\mathcal{L}^\mathrm{(T)}\right) & = \sum_b^\mathrm{(T)} \ln\left( \mathcal{P}^\mathrm{(T)}_b(T_\mathrm{EW}^\mathrm{Clu}) \right) \text{ .} \label{eq:T_LL}
                \end{align}
                Here, we compute the probability $\mathcal{P}^\mathrm{(T)}_b$ of measuring the emission-weighted temperature for the  profile bin $b$ given our model  $T_\mathrm{EW}^\mathrm{Clu}$. 
                $\mathcal{P}^\mathrm{(T)}_b$ is estimated with the normalized posterior distribution of the temperature obtained from the spectral fit shown in \Fref{SpecFit_TempProf}.
                For our maximum-likelihood estimation, we derive the posterior distributions of our model parameters and the Bayesian evidence with the nested sampling Monte Carlo algorithm \textsc{MLFriends} \citep{Buchner2014,Buchner2019} using the Python package 
                    \textsc{UltraNest}%
                    \furl{https://johannesbuchner.github.io/UltraNest}
                    (v2.2.0).

\begin{figure*}
\begin{center} 
        \resizebox{0.955\hsize}{!}{\includegraphics{\FigRootA _3x4_Prof_v3_0.png}} 
        \caption{\label{f:Fit_AllProf} 
                Comparison of our measurement (green) of the XSB profiles (\Fref{Total_XSB})
                and temperature profiles (\Fref{SpecFit_TempProf}) and their corresponding best-fit models (incl. $1\sigma$ level) for the isothermal beta-profile model (\Sref{s:Mod_IsoBeta}) in blue and for the polytropic NFW-profile model (\Sref{s:Mod_PolyNFW}) in orange.
                For visualization purposes, the thickness of the green crosses is reduced for $>2\arcmin$.
                The posterior distributions of the best-fit model parameters are shown in \Fref{Corner_FitPara_1DFit_Clu}.
                {The dashed vertical line shows the separation of the {core region} ($<100\,\kpc$) and the {without-core profile} (\Sref{s:ExCore}).}
        }
\end{center}            
\end{figure*}   

%
\begin{table*}
\begin{center}
\caption{\label{t:BestFitPara} 
        Best-fit values and $1\sigma$ levels of the free cluster parameters of both models.
    }
\begin{tabular}{c|c@{\hspace{1ex}}c@{\hspace{1ex}}c@{\hspace{1ex}}c@{\hspace{1ex}}c|c@{\hspace{1ex}}c@{\hspace{1ex}}c@{\hspace{1ex}}c@{\hspace{1ex}}c}
\hline 
           &  \multicolumn{5}{c}{Isothermal beta-profile} & \multicolumn{5}{c}{Polytropic NFW-profile} \\
    & $n_{e,\mathrm{Out}}$                     & $r_c$                  & $T_\mathrm{iso}$       & $n_{e,\mathrm{Core}}$                  & $T_\mathrm{Core}$          & $n_{e,\mathrm{Out}}$                     & $r_s$                  & $c_{\Delta}$        & $n_{e,\mathrm{Core}}$                  & $T_\mathrm{Core}$          \\
    & [$10^{-3}\,\mathrm{cm^{-3}}$] & [Mpc]                  & [keV]                  & [$10^{-3}\,\mathrm{cm^{-3}}$] & [keV]               & [$10^{-3}\,\mathrm{cm^{-3}}$] & [Mpc]                  & [-]                 & [$10^{-3}\,\mathrm{cm^{-3}}$] & [keV]               \\
\hline
 A  & $3.1_{-0.4}^{+0.5}$           & $0.18\pm{0.02}$        & $4.0\pm{0.3}$          & $4.1\pm{0.4}$                 & $3.2_{-0.7}^{+0.6}$ & $4.7_{-0.6}^{+0.7}$           & $0.46_{-0.05}^{+0.06}$ & $3.0_{-0.3}^{+0.4}$ & $5.9\pm{0.5}$                 & $3.1\pm{0.7}$       \\
 B  & $1.1\pm{0.1}$        & $0.34\pm{0.03}$        & $2.3\pm{0.2}$          & $1.2_{-0.5}^{+0.4}$           & $2.4_{-1.0}^{+1.4}$ & $1.6\pm{0.2}$                 & $1.1_{-0.1}^{+0.2}$    & $0.9_{-0.1}^{+0.2}$ & $1.4_{-0.6}^{+0.5}$           & $2.3_{-1.0}^{+1.5}$ \\
 C  & $1.0\pm{0.2}$                 & $0.28_{-0.05}^{+0.06}$ & $1.5_{-0.1}^{+0.2}$ & $<2.2$           & $0.5 - 6.0$           & $1.4_{-0.3}^{+0.4}$           & $0.9_{-0.2}^{+0.4}$    & $0.8\pm{0.3}$       & $<2.8$                 & $0.5 - 6.0$           \\
\hline
\end{tabular}
\tablefoot{
        The corresponding posterior distributions are shown in \Fref{Corner_FitPara_1DFit_Clu}.
As for cluster C the core-model parameters were not properly determined,
        we show for $n_{e,\mathrm{Core}}$ its $3\sigma$ upper limit ($\lesssim99.73\,\%$) and for $T_\mathrm{Core}$ its fit boundaries.
        We note that the $n_{e}$ parameters of both models cannot be directly compared, given the different physical assumptions of both models (see \Sref{s:Mod_Prof}).
}
\end{center}
\end{table*}            

\begin{table*}
\begin{center}
\caption{\label{t:DevPara} 
        Median and $1\sigma$ levels of $R_{500}$-related quantities.
        }
\begin{tabular}{cc|cc|cc|cc}
\hline
\multirow{2}{*}{Symbol} & \multirow{2}{*}{Unit} &  \multicolumn{2}{c}{Cluster A} & \multicolumn{2}{c}{Cluster B} & \multicolumn{2}{c}{Cluster C} \\
  &  & \multicolumn{1}{c}{Iso-$\beta$} & \multicolumn{1}{c|}{$\gamma$-NFW} & \multicolumn{1}{c}{Iso-$\beta$} & \multicolumn{1}{c|}{$\gamma$-NFW} & \multicolumn{1}{c}{Iso-$\beta$} & \multicolumn{1}{c}{$\gamma$-NFW}  \\
\hline
 $R_{500}$              & Mpc                           & $0.85\pm{0.03}$                   & $0.86\pm{0.03}$        & $0.56\pm{0.03}$        & $0.55_{-0.03}^{+0.04}$ & \multicolumn{2}{c}{$0.44_{-0.05}^{+0.04}$}                         \\
 $\theta_{500}$         & arcmin                        & \multicolumn{2}{c|}{$2.7\pm{0.1}$}  & \multicolumn{2}{c|}{$1.8\pm{0.1}$} & $1.4_{-0.2}^{+0.1}$                     & $1.4\pm{0.1}$          \\
 $M_{t,500}$            & $10^{14}\,\mathrm{M_{\odot}}$ & $2.4_{-0.2}^{+0.3}$               & $2.5\pm{0.3}$          & \multicolumn{2}{c|}{$0.7\pm{0.1}$} & \multicolumn{2}{c}{$0.3\pm{0.1}$} \\
 $M_{\mathrm{gas},500}$ & $10^{13}\,\mathrm{M_{\odot}}$ & \multicolumn{2}{c|}{$2.2\pm{0.1}$} & $1.0\pm{0.1}$ & $0.9\pm{0.1}$ & $0.5\pm{0.1}$ & $0.4\pm{0.1}$                                       \\
\hline
\end{tabular}
\tablefoot{
    Values are derived from our best-fit models of the isothermal beta-profile model (``Iso-$\beta$'') and of the polytropic NFW-profile model (``$\gamma$-NFW'').
        The same quantities for $R_{200}$ are shown in \Tref{t:DevPara_r200}. Here,
        $\theta$ is the angular scale,
        $M_t$ the total hydrostatic mass, and
        $M_\mathrm{gas}$ the gas mass of the corresponding physical size $R$. 
        Within the same row, values are only shown once if they are the same for both models.
}
\end{center}
\end{table*}

\begin{table*}
\begin{center}
\caption{\label{t:LumPara} 
        Median and $1\sigma$ levels of the luminosity in the $0.5-2.0\,\mathrm{keV}$ band for different apertures.
        }
\begin{tabular}{c|cc|cc|cc}
\hline
\multicolumn{7}{c}{Luminosity [$10^{43}\,\mathrm{erg\,s^{-1}}$] ($0.5-2.0\,\mathrm{keV}$)} \\
\hline
\multirow{2}{*}{Aperture} & \multicolumn{2}{c}{Cluster A} & \multicolumn{2}{c}{Cluster B} & \multicolumn{2}{c}{Cluster C} \\
   & \multicolumn{1}{c}{Iso-$\beta$} & \multicolumn{1}{c|}{$\gamma$-NFW} & \multicolumn{1}{c}{Iso-$\beta$} & \multicolumn{1}{c|}{$\gamma$-NFW} & \multicolumn{1}{c}{Iso-$\beta$} & \multicolumn{1}{c}{$\gamma$-NFW}  \\
\hline
$<R_{500}$                    & \multicolumn{2}{c|}{$6.5\pm{0.3}$} & $3.0\pm{0.2}$   & $2.9\pm{0.2}$          & \multicolumn{2}{c}{$1.4\pm{0.2}$}                        \\
$R_\mathrm{Core} - R_{500}$   & \multicolumn{2}{c|}{$5.2\pm{0.3}$} & $2.7\pm{0.2}$   & $2.6\pm{0.2}$          & \multicolumn{2}{c}{$1.2\pm{0.2}$}                        \\
$<0.3\,\Mpc$                  & \multicolumn{2}{c|}{$4.1\pm{0.1}$} & $1.6\pm{0.1}$ & $1.5\pm{0.1}$        & \multicolumn{2}{c}{$1.0\pm{0.1}$} \\
\hline
\end{tabular}
\tablefoot{
    Values are derived from our best-fit models of the isothermal beta-profile model (``Iso-$\beta$'') and of the polytropic NFW-profile model (``$\gamma$-NFW'').
        Within the same row, values are only shown once if they are the same for both models.
        $R_\mathrm{Core}$ equals $0.1\,\Mpc\sim[0.1,0.2,0.2] \times R_{500}$ for cluster [A,B,C], respectively.
        The estimates for $<0.3\,\Mpc$ are used for the comparison with the temperature-luminosity scaling relation (\Sref{s:ScalRela}).
}
\end{center}
\end{table*} 

\begin{table*}
\begin{center}
\caption{\label{t:TempPara} 
        Median and $1\sigma$ levels of the temperature [keV] for different aperture.
        }
\begin{tabular}{c|cc|cc|cc}
\hline
\multicolumn{7}{c}{Temperature [keV]} \\ 
\hline
\multirow{2}{*}{Aperture} & \multicolumn{2}{c}{Cluster A} & \multicolumn{2}{c}{Cluster B} & \multicolumn{2}{c}{Cluster C} \\
   & \multicolumn{1}{c}{Iso-$\beta$} & \multicolumn{1}{c|}{$\gamma$-NFW} & \multicolumn{1}{c}{Iso-$\beta$} & \multicolumn{1}{c|}{$\gamma$-NFW} & \multicolumn{1}{c}{Iso-$\beta$} & \multicolumn{1}{c}{$\gamma$-NFW}  \\
\hline
 $<R_{500}$                           & $4.0_{-0.2}^{+0.3}$ & $3.8\pm{0.2}$       & \multirow{3}{*}{$2.3\pm{0.2}$} & $2.3_{-0.1}^{+0.2}$ & $1.6_{-0.2}^{+0.3}$    & $1.6_{-0.2}^{+0.3}$    \\
 $<0.3\,\Mpc$                         & $3.9_{-0.2}^{+0.3}$ & $4.0_{-0.2}^{+0.3}$ &        & $2.3\pm{0.2}$       & $1.6_{-0.2}^{+0.4}$    & $1.6_{-0.2}^{+0.4}$    \\
 $R_\mathrm{Core} - R_\mathrm{Spec.}$ & $4.0\pm{0.3}$       & $3.9\pm{0.3}$       &          & $2.3_{-0.1}^{+0.2}$ & $1.5_{-0.1}^{+0.2}$ & $1.5\pm{0.1}$ \\
\hline
\end{tabular}
\tablefoot{
    Values derived from our best-fit models of the isothermal beta-profile model (``Iso-$\beta$'') and of the polytropic NFW-profile model (``$\gamma$-NFW'').
    $R_\mathrm{Core}$ equals $0.1\,\Mpc\sim[0.1,0.2,0.2] \times R_{500}$ for clusters [A,B,C], respectively.
        Excluding the cluster core does not alter the temperature estimates significantly.
        Within one column, values are only shown once if more than two consecutive rows have the same value.
        The estimates for $<0.3\,\Mpc$ are used for the comparison with the temperature-luminosity scaling relation (\Sref{s:ScalRela}).
        The $R_\mathrm{Core} - R_\mathrm{Spec.}$ aperture uses the same radial limit as the broad profile range, 
        which was used in the X-ray spectroscopy to estimate average cluster properties (\Sref{s:Clu_Ave}).
        Comparing the estimates from the best-fit models and the spectroscopy for this aperture reveals that they are
        consistent with each other (see \Tref{t:SpecFit_general}).
}
\end{center}
\end{table*} 

        \subsection{Results} \label{s:results} 
                \subsubsection{Best-fit models and derived quantities} \label{s:1DXfit}  \label{s:DerPara}
                In \Fref{Fit_AllProf}, we compare our best-fit profile models with our data.
                This illustrates that both best-fit models, that is, the isothermal beta-profile and the polytropic NFW-profile model, can describe the measured profiles of all clusters rather well.
                In \Tref{t:BestFitPara}, we list the best-fit values of the cluster parameters of each model
                and in \Fref{Corner_FitPara_1DFit_Clu} we show their corresponding posterior distributions.
                We should note that the free cluster parameters do not show any strong degeneracy with the free background-model parameters described in \Sref{s:1D_XSB_model_tot}.
                
                For cluster C, the core-model parameters were not properly determined because of a lack of sufficient observational constraints,
                making the results dependent on their fit boundaries: $[0.4,10.0]\times10^{-3}\,\mathrm{cm^{-3}}$ for $n_{e,\mathrm{Core}}$ and $[0.5,6.0]\,\mathrm{keV}$ for $T_\mathrm{Core}$.
                However, those boundaries 
                were chosen to be wide enough to encompass the {expected range for clusters}. 
        
                We use both best-fit models, that is, the isothermal beta-profile and the polytropic NFW-profile model, to derive important physical quantities, such as size, mass, luminosity, and average temperature for different apertures.
                Unsurprisingly, both models give consistent estimates.
                In \Tref{t:DevPara}, we list $R_{500}$
                and its associated angular scale ($\theta$), total hydrostatic mass ($M_t$), and gas mass ($M_\mathrm{gas}$). 
                {The same quantities for $R_{200}$ are shown in \Tref{t:DevPara_r200}.
                This shows that cluster A is more than twice as massive as both other clusters combined based on their $R_{500}$ values.
                {The combined $R_{200}$ mass of all three cluster remains below $\sim10^{15}\,\mathrm{M_{\odot}}$ (\Tref{t:DevPara_r200}).}
                The $R_{500}$ regions are not overlapping with each other (see \Fref{Maps_Masks})
                but the $R_{200}$ region of cluster A has a small overlap with the $R_{200}$ regions of the other clusters in the projected plan.}
                
                In \Tref{t:LumPara}, we list the derived luminosities ($L$) for different apertures, 
                which show that including or excluding the cluster core makes a significant {difference}, especially for cluster A.
                In \Tref{t:TempPara}, we list the derived temperature for different apertures.
                The estimates for the $<0.3\,\Mpc$ {region} are used for the comparison with a temperature--luminosity scaling relation (\Sref{s:ScalRela}).
                The $R_\mathrm{Core} - R_\mathrm{Spec.}$ aperture uses the same radial limit as the broad profile range, 
                which was {used to} estimate average cluster properties (\Sref{s:Clu_Ave}).
                The values from this direct measurement (\Tref{t:SpecFit_general}) and the best-fit models are consistent with each other.         
                
\begin{figure*}
\begin{center}
        \resizebox{\hsize}{!}{\includegraphics{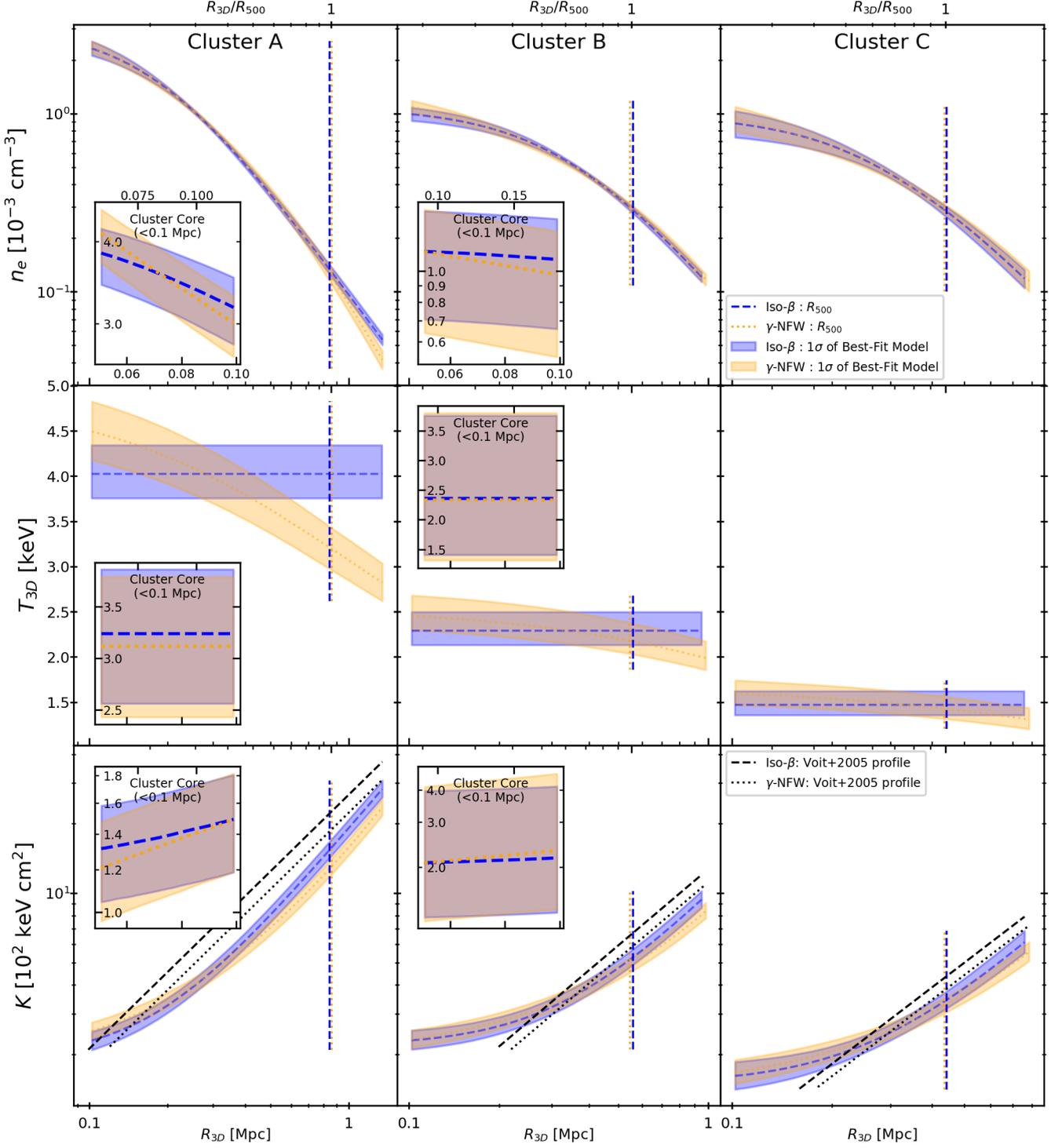}} 
        \caption{\label{f:EntroProf} 
        \label{f:3D_prof_A} 
        Electron number density, temperature, and entropy profiles of the best-fit model up to $R_{200}$.
        Blue areas and dash curves show the isothermal beta-profile model (\Sref{s:Mod_IsoBeta}).
        Orange areas and dotted curves show the polytropic NFW-profile model (\Sref{s:Mod_PolyNFW}).
        Curves show the median and areas show the $1\sigma$ level of the models.
        {The main plots show the without-core profiles ($\geq 100\,\kpc$) and the plot inserts show the core region ($<100\,\kpc$).
        This visual separation is necessary because our model can create nonphysical discontinuities between both regions, especially for cool-core clusters (\Sref{s:ExCore}).
        For each cluster, all inserts have the same x-axis range but for clarity the tick labels are only shown in the top row.
        The core region of cluster C is not shown because its core-region parameters were not properly determined (\Tref{t:BestFitPara}).
        For the entropy profiles,}
        the black lines show the self-similar prediction by \citet{Voit2005}, $K(r_\mathrm{3D}) = 1.32 \, K_{200} \, (r_\mathrm{3D}/R_{200})^{1.1}$.
        }
\end{center}            
\end{figure*}    
                
                \subsubsection{Entropy and the cluster core}  
                \label{s:core}  \label{s:entropy}       
                With the best-fit model of the 3D gas density and temperature profiles, we can derive the X-ray entropy profile as follows:
                \begin{align} 
                        K & = T \, \left( n_e \right)^{-2/3}  \label{eq:K_X} \text{ .}
                \end{align}
                All three profiles are shown for each cluster in \Fref{EntroProf}.
                {The core region ($<100\,\kpc$) is only shown in a plot insert to visually separate it from the without-core profile ($\geq 100\,\kpc$) because our model {can induce} nonphysical discontinuities between both regions, especially for cool-core clusters (\Sref{s:ExCore}).
                For cluster C, {the} core region is not shown because {its parameters} were not properly determined (\Tref{t:BestFitPara}).}
                
                For the entropy profiles in \Fref{EntroProf}, we also show  a self-similar model by \citet{Voit2005} using black lines, namely
                $K(r_\mathrm{3D}) = 1.32 \, K_{200} \, (r_\mathrm{3D}/R_{200})^{1.1}$,
                which does not take stellar or active galactic nucleus (AGN) feedback into account.
                For cluster A, the entropy profile follows this model on most scales rather well (apart from the offset in the normalisation).
                Moreover, its core appears to be {denser} and cooler {than} the without-core profile, 
                which can be deduced from the best-fit values of its isothermal beta-profile model (\Tref{t:BestFitPara}).
                These observations suggest that cluster A most likely falls into the class of {relaxed cool-core} clusters.
                For cluster B, the entropy profile flattens around $\sim0.5\,R_{500}$
                suggesting an entropy excess towards the cluster center with respect to the self-similar model.
                Moreover, there is no significant discontinuity between core region and the without-core profile,
                suggesting that the core is {neither denser} nor cooler {than} the without-core profile.
                These observations indicate that cluster B falls rather into the class of {disturbed noncool-core} clusters.
                Our tentative classification for clusters A and B is also supported by a comparison with previous entropy measurements with large cluster samples \citep{Cavagnolo2009,Ghirardini2017}.
                For cluster C, it appears that the entropy profile flattens around $\sim0.5\,R_{500}$ 
                but due to the unconstrained core region, we cannot {assess} how density and temperature {change} towards the cluster center
                and therefore {we} refrain from making any classification.


\section{Discussion} \label{s:Dis}

\begin{figure} 
\begin{center} 
        \resizebox{\hsize}{!}{\includegraphics{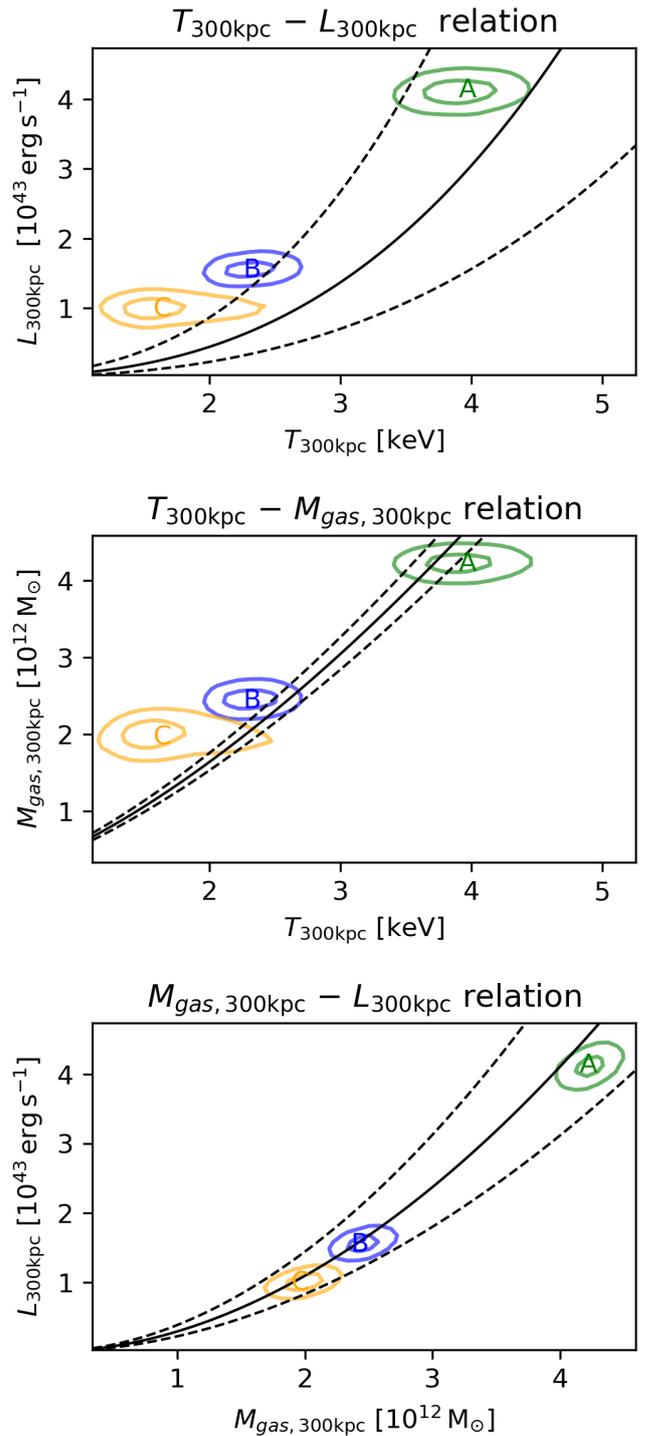}} 
        \caption{\label{f:ScaRel} 
                Comparison of derived cluster parameters with scaling relations derived from the 100 brightest clusters
{of the \xxl{} sample} \citep{2019Sereno}.
                The luminosity is given for the $0.5-2.0\,\mathrm{keV}$ band and the aperture for all quantities is $R<0.3\,\Mpc$.
                The black solid and dashed curves show the median and intrinsic scatter of the scaling relations, respectively.
                Green, blue, and orange contours correspond to clusters A, B, and C, respectively.
                The contours show the $1\sigma$ and $2\sigma$ levels
                {of the isothermal beta-profile model.
                As the polytropic NFW-profile model has almost the same contours, they are omitted for clarity.
                }
        }
\end{center}            
\end{figure}    

        \subsection{Cluster properties in respect to scaling relations} \label{s:ScalRela}
        In \Fref{ScaRel}, we {compare} the temperature $T$, luminosity $L$ ($0.5-2.0\,\mathrm{keV}$), and gas mass $M_\mathrm{gas}$ derived from our best-fit models
        with three scaling relations derived from the 100 brightest clusters {of the \xxl{} sample} \citep{2019Sereno}.
    The redshift {and mass of our {clusters} fall} well within the covered range of this sample.
        All clusters follow the $M_\mathrm{gas}-L$ relation relatively well.
        For the $T-L$ and $T-M_\mathrm{gas}$ relations, cluster C shows the strongest {disagreement}.
        However, when compared with the {actual} measurements for the \xxl{} sample \citep[fig.~2 of][]{2019Sereno},
        cluster C does not appear as an outlier in this sample.
        This indicates that all clusters of our system have consistent properties with respect to the \xxl{} 100 brightest clusters. 
        {As about $80\,\%$ of the clusters in the sample are identified as single clusters, 
        the consistency with them may indicate that our clusters have similar properties to typical single clusters.}

\begin{figure} 
\begin{center} 
        \resizebox{\hsize}{!}{\includegraphics{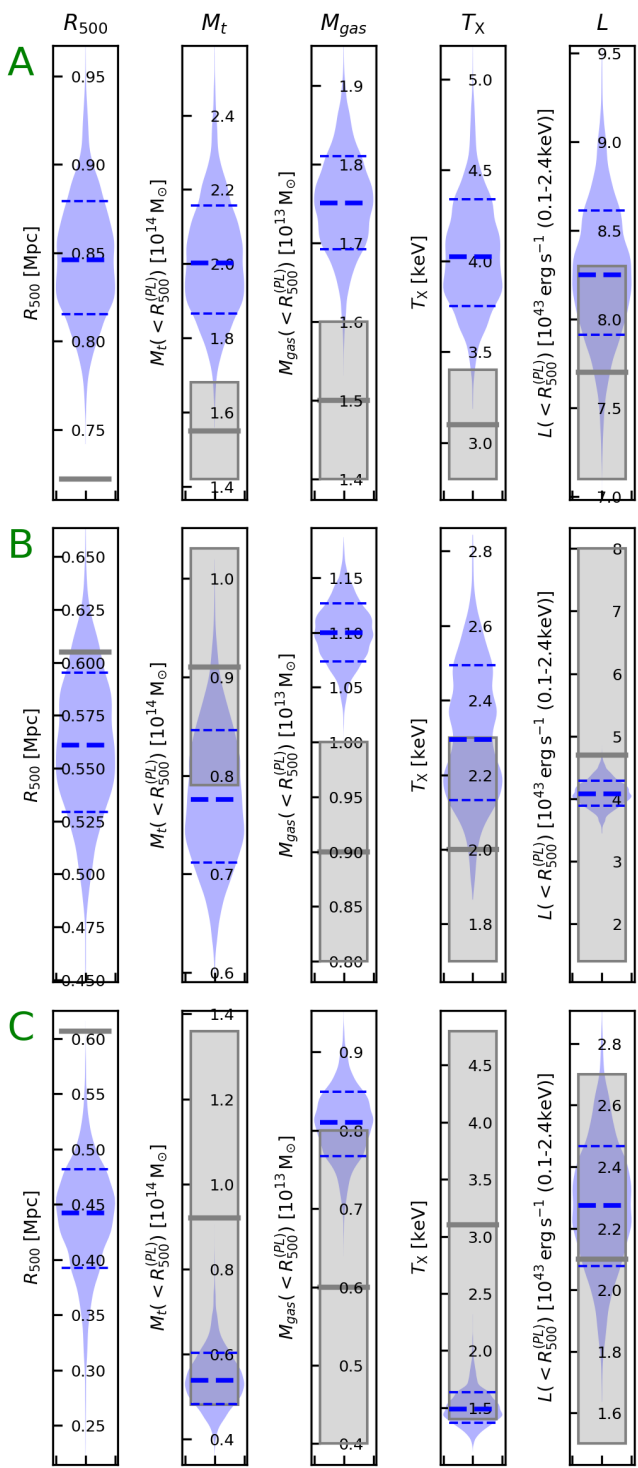}} 
        \caption{\label{f:DevPara_ModelComp_1DFit} 
                Comparison of cluster parameters derived from our best fit {of the  isothermal beta-profile model in blue} and estimated by \PlXI\ (table~2) in gray.
                {As the results for the polytropic NFW-profile model are almost the same, they are omitted for clarity.}
                The blue thick and thin dash lines are the median and $1\sigma$ levels of our model.
                The gray areas show the $1\sigma$ levels of \PlXI{}, except for $R_{500}$, which was not provided. 
                Top, middle, and bottom rows correspond to clusters A, B, and C, respectively.
        See \Sref{s:DerPara} for more details.
        }
\end{center}            
\end{figure}    

        \subsection{Measurement comparison} \label{s:PL_Meas}
        In \Fref{DevPara_ModelComp_1DFit},
        we compare our estimates for $R_{500}$, $M_{t,500}$, $M_{gas,500}$, $T_\mathrm{X}$, and $L_{500}$ with those from \PlXI{}, which only used the shallow observation for their analysis.
        There, we can see that their $R_{500}$ estimate is {significantly} different {from} ours, which reflects {the different approaches} in estimating $R_{500}$.
        To have a fair comparison of $R_{500}$-dependent quantities, we {use} their {measurement of $R_{500}$, symbolized as $R^{(PL)}_{500}$}.
        Moreover, 
        {for increased consistency with the measurements of} \PlXI{}, $T_\mathrm{X}$ was estimated for $(0.15-0.75)\,R^{(PL)}_{500}$ and the luminosity $L(<R^{(PL)}_{500})$ was estimated for the $0.1-2.4\,\mathrm{keV}$ band {from our best-fit models}.
        
        In respect to the $R_{500}$-dependent quantities, 
        we can see in \Fref{DevPara_ModelComp_1DFit} that for clusters B and C the measurements of \PlXI{} are consistent with ours,
        {except for $M_{gas}$ of cluster B.}
    {We note that \PlXI{} state that they did not account for systematic uncertainties related to redshift uncertainties or high background levels, 
    which likely means their errors are underestimated.}
        {For instance,} their uncertainties for cluster A appear to be of similar size to ours, while they only used the shallow observation, which has an approximately four times lower exposure time than the deep observation.
        Even taking this into account
        {cannot explain the discrepancy for the estimates of $M_{t,500}$, $M_{gas,500}$, and $T_\mathrm{X}$ for cluster A.}
    {The {difference} might be due to the choice of approach in the spatial and spectral modeling of the cluster emission.}
        Most notably,
        \PlXI{} subtracted the IBKG beforehand and left the power-law index of the extragalactic CXB model free
        and used the \texttt{MEKAL} model for the cluster emission. 
        
    \subsection{Interactions between clusters} \label{s:cint} 
    One way to search for signs of interaction between clusters is to detect enhanced (or irregular) X-ray emission between clusters \citeg{Planck_Int_VI_2013}.
    For {the present triple system}, this requires modeling the emission of all clusters simultaneously,
    because they all have similar angular {distances and}
    are distributed in the projected {plane} almost like an equilateral triangle with an edge length of $\sim2.0\,\Mpc$ ($\sim6\arcmin$),
    {which does not rule out the scenario that all three clusters have a significant interaction with each other at the same time.}
    However, when comparing our best-fit model to a smoothed count-rate image for each detector,
    no significant deviation from the total signal of all clusters, the CXB, and the IBKG was detected.   
    This could mean that there has not been any direct cluster interaction so far,
    which would not be unexpected because none of the clusters have overlapping $R_{500}$ regions.
    This would also be in line with the finding that all clusters are {consistent} with the scaling relation of the \xxl{} 100 brightest clusters,
    where $80\,\%$ of them are not associated with a multi-cluster system (\Sref{s:ScalRela}).     
    However, our X-ray data alone are not sensitive enough to be used to measure such an excess signal,
    because the XSB of each cluster drops below the CXB already within its $R_{500}$ region.
    Hence, a joint analysis of X-ray and tSZ data would be required to increase the sensitivity and potentially detect signs of interactions between clusters, because the tSZ effect is more sensitive to the emission in the cluster outskirts \citeg{Eckert2016a}.
    The latter makes such a joint analysis particularly suitable for revealing the presence and the properties of WHIM within the system.    

    \subsection{Comparison of \planck{}-detected triplet systems} \label{s:CompSC}
    The {triple system (hereafter TS1)} analyzed by \citet{Planck_Int_VI_2013} (\texttt{PLCK G214.6+36.9}) and that analyzed in this work (hereafter TS2)
    are the only two known \planck{}-detected triplet-cluster systems to date.
    Nevertheless, both cases might only be  apparent triplet systems, because it remains unclear as to whether or not all members are at the same redshift.
    TS1 is more distant than TS2 to the observer, with redshifts of $z \approx 0.45$ and $z \approx 0.37$, respectively.
TS1 is also more massive with about twice as much accumulative $R_{500}$ mass,
    which is almost evenly distributed among the SC members with a proportion of $\sim1:1.1:1.4$ for clusters A, B, and C, respectively,
    while for TS2 the proportion is $\sim7:2:1$.
    For TS2, one cluster is more than twice as massive as both other clusters combined.
    Both {systems} contain one relaxed cluster with a cool core (cluster A in both cases)
    while the other clusters appear to be more disturbed and without a cool core (or remain without classification), based on the derived gas density, temperature, and entropy profiles from X-ray data.
    It is interesting to note that for TS1 the relaxed {cool}-core cluster is the least massive, while for TS2 it is the most massive.
    
    The clusters of TS1 are distributed in the projected {plane} almost like an isosceles triangle with cluster C at the top and at a distance of $\sim2.5\,\Mpc$  from the other clusters, which are both separated by only $\sim1.1\,\Mpc$.
    The clusters of TS2 have more similar angular distances between each other and 
    are distributed in the projected {plane} almost like an equilateral triangle with an edge length of $\sim2.0\,\Mpc$. 
    For TS1, the two clusters closest to each other have overlapping $R_{500}$ regions in the projected {plane} but no enhanced X-ray emission ---an indication of interaction--- was detected between them.
    For TS2, none of the clusters have overlapping $R_{500}$ regions and no enhanced X-ray emission was detected between them.
    
    This comparison shows that {none of the systems show} signs of cluster interactions
    and that they have quite different configurations based on their X-ray data.
    This suggests rather different cluster merger scenarios for both SCs, under the assumption that all clusters of each SC are at 
    about the same redshift, {and are indeed merging systems.}

\section{Summary} 
\label{s:sum}  \label{s:con} 
Multi-cluster systems are  an important structure formation probe in the Universe.
{Two triplet-cluster systems have been discovered from the follow-up campaign of \planck{}-detected clusters \citep{PlanckEarlyVIII2011,PlanckEarlyIX2011}  with \xmm.}
For one system, \texttt{PLCK G214.6+36.9}, a multi-wavelength analysis has already been conducted \citep{Planck_Int_VI_2013}.
In the present work, we study the X-ray emission observed by \xmm\ of the other system, \texttt{PLCK G334.8-38.0}, 
which represents the first step in a multi-wavelength study.

Our X-ray {analysis} reveals that the system is located at $z=0.37\pm{0.01}$ (\Sref{s:z_spec}).
Although, our measurement is not precise enough to confirm that all three clusters are part of the same system,
meaning that a subsequent study with optical spectroscopy is required for verification.
The X-ray {analysis} also provides a temperature profile for each cluster (\Sref{s:Tprof})
and supports the assumption of $0.3$  times solar metallicity for their ICM (\Sref{s:M_spec}).

We simultaneously fitted the spectroscopic temperature profile (\Fref{SpecFit_TempProf}) with the XSB profiles of all three \xmm{} detectors (\Fref{Total_XSB}) in order to constrain the physical properties of each cluster (\Sref{s:mod}).
This revealed a hydrostatic mass of $\sim [2.5,0.7,0.3] \times 10^{14}\,\mathrm{M_{\odot}}$
and an average temperature of $\sim [3.9,2.3,1.6]\,\mathrm{keV}$) within $R_{500}$ for clusters A, B, and C, respectively (\Sref{s:DerPara}).
Hence, cluster A is more than twice as massive as both other clusters combined, showing an uneven distribution of mass within the system, whose total mass is below $\sim10^{15}\,\mathrm{M_{\odot}}$ based on the $M_{200}$ mass of all clusters.

With our best-fit model, we derive the X-ray entropy profile for each cluster (\Sref{s:entropy}).
This suggests that the brightest/most massive cluster A appears to be a relaxed cool-core cluster, 
which is also supported by the temperature decrease and gas density increase towards its center (\Fref{3D_prof_A}).
The second brightest and second-most massive cluster
(B) appears to be a disturbed noncool-core cluster
and the X-ray signal of the {third} cluster (C) was too weak to make such a classification.

No sign of cluster interaction was found when searching for enhanced X-ray emission between the clusters,
which is not unexpected, because none of the clusters have overlapping $R_{500}$ regions (\Sref{s:cint}).
This is also in line with the consistency of cluster properties with scaling relations based on single clusters (\Sref{s:ScalRela}).
However, our X-ray data alone {do not permit us to detect} significant cluster emission beyond $R_{500}$.
Hence, a joint analysis of X-ray and tSZ is {required,
which} may also reveal the presence of WHIM within the system.

Comparison of the two \planck{}-detected triplet-cluster systems reveals that neither shows signs of cluster interactions
and that the two have quite different configurations based on their X-ray data alone (\Sref{s:CompSC}).
This suggests rather different cluster merger scenarios, under the assumption that all clusters of each system are at about the same redshift.

\begin{acknowledgements}
        The authors acknowledge helpful discussion with F. Gastaldello, S. Molendi,
        and the members of the ByoPiC project\furl{https://byopic.eu/team}. 
        We thank the referee for the useful comments.
        %
        This research has been supported by the funding for the ByoPiC project from the European Research Council (ERC) under the European Union's Horizon 2020 research and innovation program grant agreement ERC-2015-AdG 695561 and CNES.
        %
        This publication used observations obtained 
        %
        with \xmm{}\furl{https://www.cosmos.esa.int/web/xmm-newton}, 
        an ESA science mission with instruments and contributions directly funded by ESA Member States and NASA. 
        %
        This research has made use of data obtained from the 3XMM \xmm\ serendipitous source catalog compiled by the 10 institutes of the \xmm\ Survey Science Centre selected by ESA.
        %
        This research has made also use of  
        the application package \textsc{CIAO} (v4.11), provided by the \chandra{} X-ray Center (CXC),
        and the software packages \textsc{FTOOLS}\furl{https://heasarc.gsfc.nasa.gov/docs/software/ftools} \citep[v6.26,][]{FTOOLS}
        the Python environment \textsc{IPython}\furl{https://ipython.org} \citep{IPython} 
        and \textsc{Jupyter notebook}\furl{https://jupyter.org},
        the community-developed Python packages of \textsc{AstroPy}\furl{https://www.astropy.org} \citep{astropy:2013, astropy:2018},
        core packages of \textsc{SciPy}\furl{https://www.scipy.org/} \citep[a Python-based ecosystem of open-source software,][]{SciPy},
        such as \textsc{NumPy}\furl{https://www.numpy.org} \citep{Numpy01,Numpy02}, and \textsc{MatPlotLib}\furl{https://matplotlib.org} \citep{Matplotlib},
        and the Python package \textsc{corner}\furl{https://github.com/dfm/corner.py} \citep{corner}.
\end{acknowledgements}

\bibliographystyle{aa}

\begin{appendix}
        
\section{Energy spectrum fit of the CXB} \label{a:CXB_spec}
\label{a:IBKG_spec} 

\begin{figure*}
\begin{center} 
        \resizebox{\hsize}{!}{\includegraphics{\FigDir 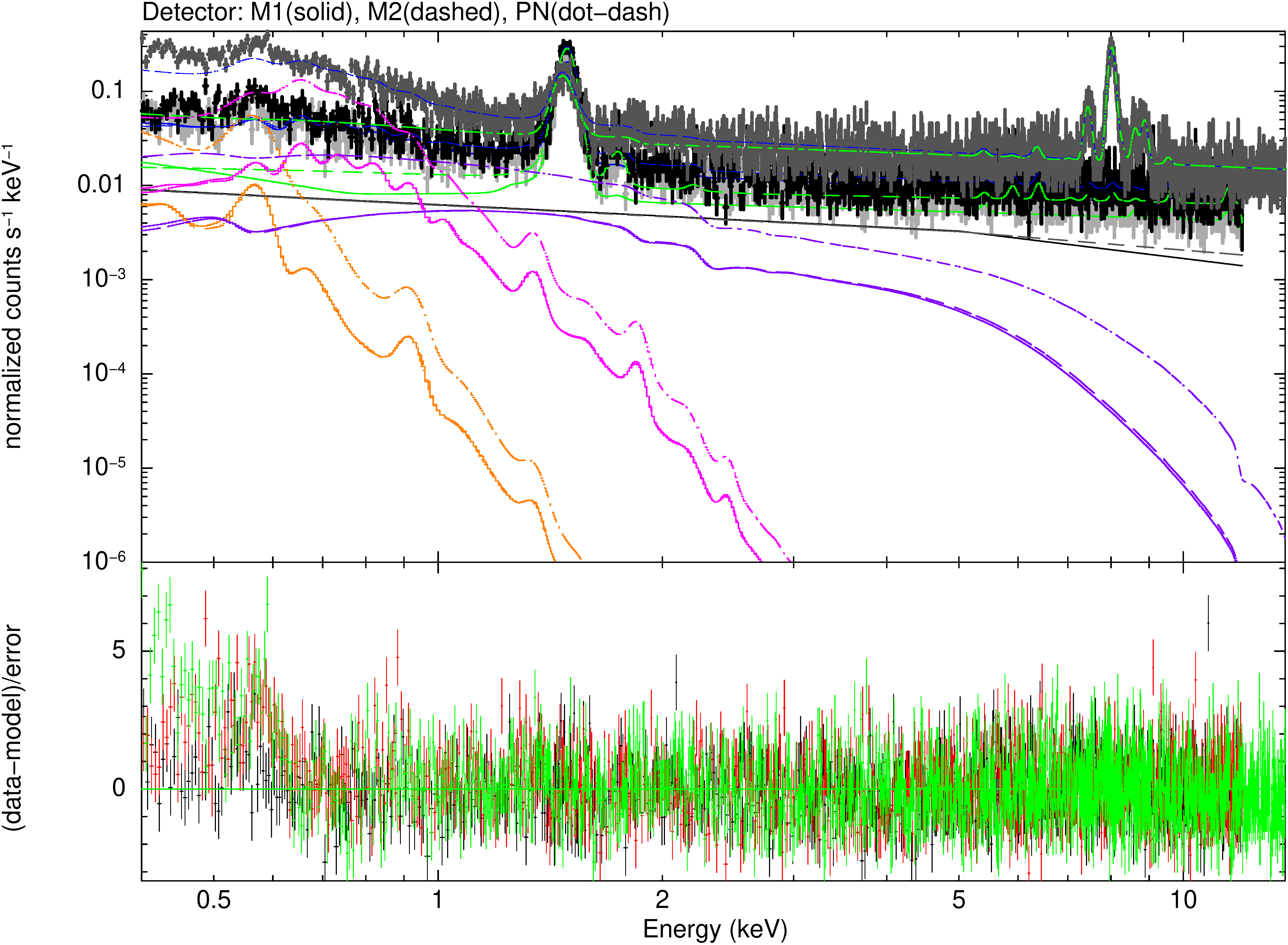}} 
        \caption{\label{f:CXB_spec_full}
                Energy spectrum of the CXB region with the best-fit model of all \xmm\ detectors of the deep observation.
                The data points of EMOS1, EMOS2, and EPN are light gray, black, and dark gray in the upper panel, and black, red, and green in the lower panel, respectively.
                The model curves of EMOS1, EMOS2, and EPN are solid, dashed, and dot-dashed, respectively (mostly overlapping for the EMOS detectors).
                For each detector, 
                the total model is shown as a blue curve, and                the models of the CXB components, namely the local hot bubble, the Galactic halo, and the extragalactic emission are shown as orange, magenta, and violet curves, respectively. 
                The [HEB,IBKG] and SPB models are shown as green and black curves, respectively (\Sref{s:SpecFit}).
        }       
\end{center}    
\end{figure*}           

\begin{figure*}
\begin{center} 
        \resizebox{\hsize}{!}{\includegraphics{\FigDir 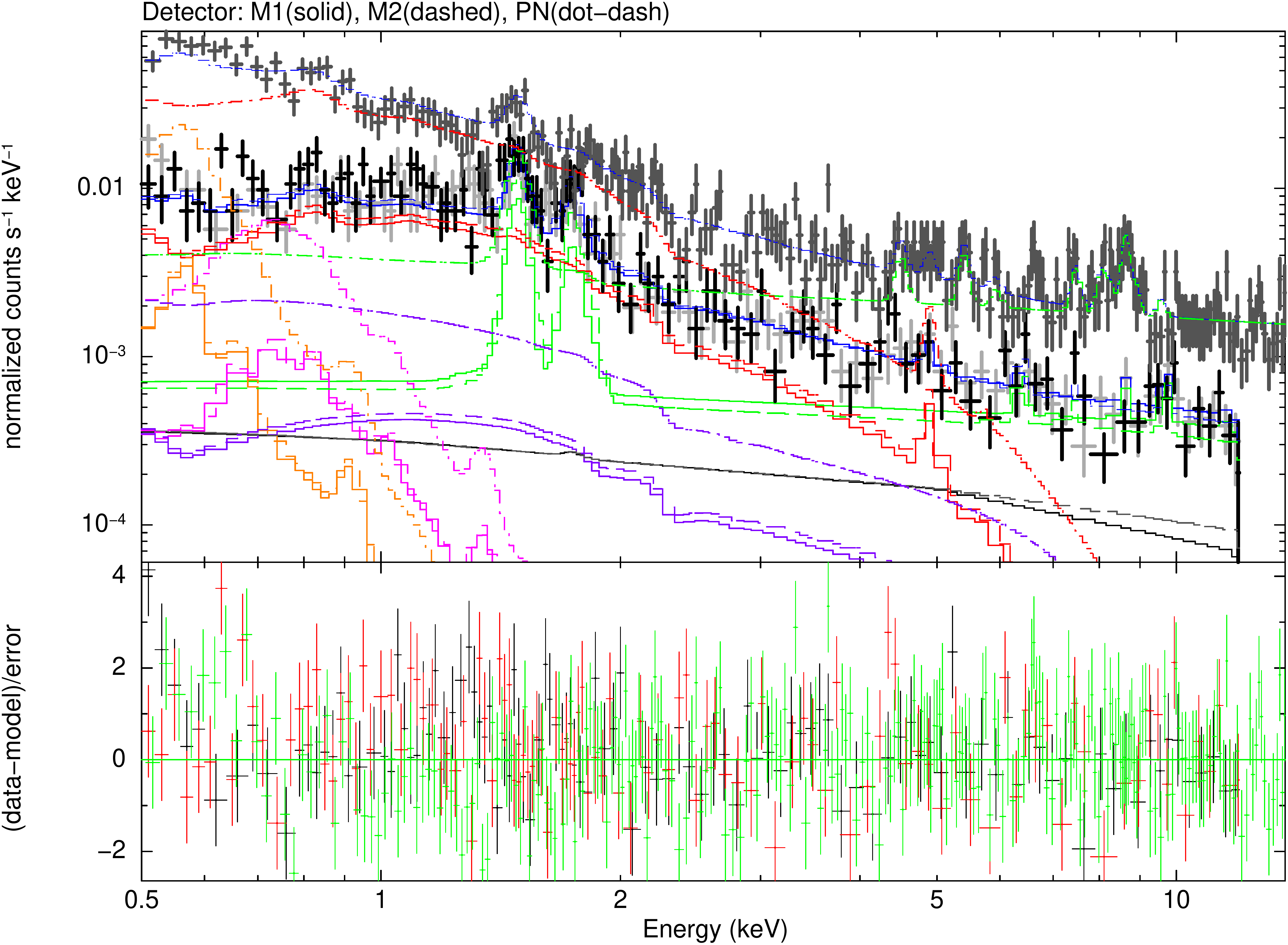}} 
        \caption{\label{f:ClA_spec_full}
                Energy spectrum of cluster A
                for its profile region $19\arcsec - 2.0\arcmin \approx (0.1 - 0.7)\times R_{500}$
                with the best-fit model of all \xmm\ detectors of the deep observation.
The same format is used here as for \Fref{CXB_spec_full};
                additionally a cluster model is shown in red for each detector.
                Here, our best-fit CXB model (\Tref{t:CXBFit}) is used as a fixed model to determine the cluster model, as described in \Sref{s:SpecFit}.
        }       
\end{center}    
\end{figure*}

\begin{table}
\begin{center}
\caption{\label{t:CXBFit} 
        Best-fit values and their $1\sigma$ uncertainties of the free CXB model parameters.
        }
\begin{tabular}{cl}
\hline
 LHB $T$     & $0.13_{-0.03}^{+0.02}$ keV         \\
 LHB $K$     & $-2.2_{-0.2}^{+0.3}$ $\,\log_{10}\left(\mathrm{deg}^{-2}\right)$ \\
 GaH $T$     & $0.29_{-0.03}^{+0.04}$ keV         \\
 GaH $K$     & $-2.64_{-0.16}^{+0.12}$ $\,\log_{10}\left(\mathrm{deg}^{-2}\right)$ \\
 ExG $K$     & $-2.80\pm{0.03}$ $\,\log_{10}\left(\mathrm{deg}^{-2}\right)$ \\
\hline
\end{tabular}
\tablefoot{
    Values obtained by modeling the X-ray energy spectrum of the CXB region (see \Fref{CXB_spec_full} and \Sref{s:SpecFit}).
    The model components are the local hot bubble (LHB), the Galactic halo (GaH), and the extragalactic emission (ExG).
        Normalization parameters ($K$) are shown in decadic logarithm.
        The corresponding values of the XSB are shown in \Fref{SpecFit_CXB_XSB}.
}
\end{center}
\end{table}     

\begin{figure*}
\begin{center}
        \resizebox{\hsize}{!}{\includegraphics{\FigDir 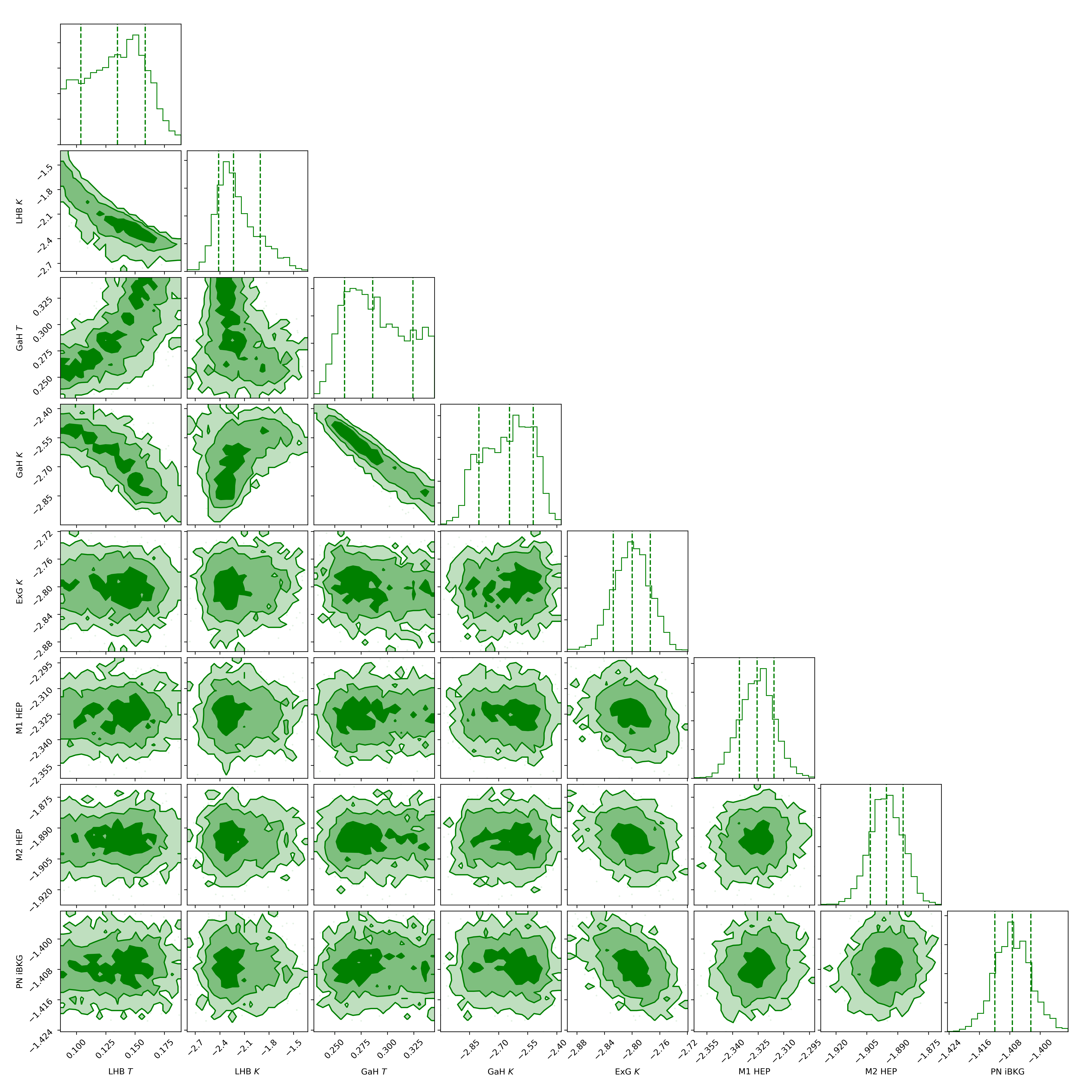}} 
        \caption{\label{f:SpecFit_CXB} 
                Posterior distribution of the free model parameters of each of the CXB components, namely the local hot bubble (LHB), the Galactic halo (GaH), and  the extragalactic emission (ExG), and the [HEB,IBKG] normalization of each \xmm\ detector after modeling the X-ray energy spectrum of the CXB region (see \Fref{CXB_spec_full} and \Sref{s:SpecFit}).
                For the 2D histograms, contours show the $1\sigma$, $2\sigma,$ and $3\sigma$ levels.
                For the 1D histograms, dashed lines show the median and $1\sigma$ levels.
                Temperature parameters ($T$) are shown in units of keV
                and normalization parameters (\eg\ $K$) are shown in decadic logarithm and units of $\log_{10}(\mathrm{deg}^{-2})$.          
                The corresponding values of the XSB for the CXB components are shown in \Fref{SpecFit_CXB_XSB}.
                }
\end{center}            
\end{figure*}   

\begin{figure}
\begin{center} 
        \resizebox{\hsize}{!}{\includegraphics{\FigDir 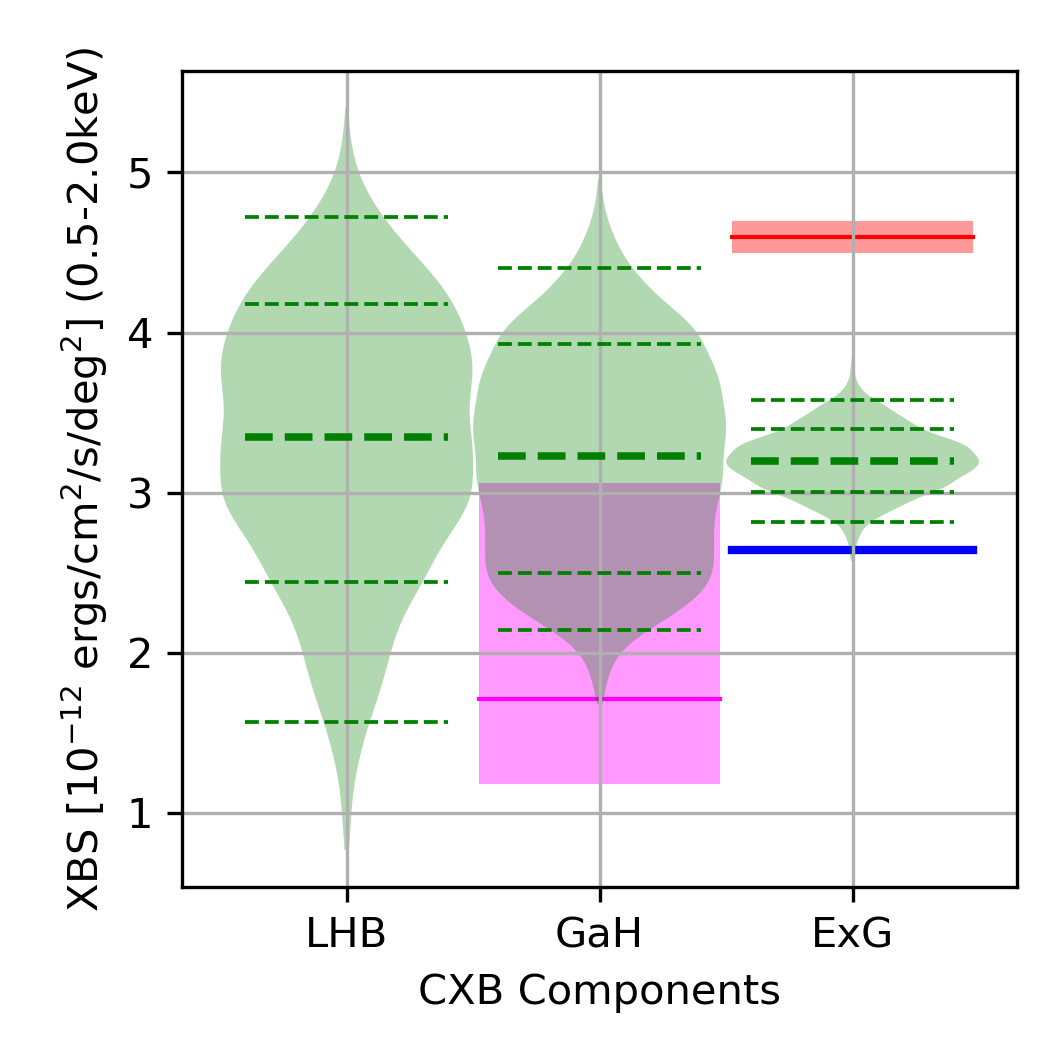}} 
        \caption{\label{f:SpecFit_CXB_XSB} 
                XSB of the CXB components, the local hot bubble (LHB), the Galactic halo (GaH), and  the extragalactic emission (ExG),
                derived from the X-ray energy spectrum fit of the CXB region (\Fref{SpecFit_CXB}).
                Thick and thin dashed lines are median and $1\sigma$ and $2\sigma$ levels.
                For GaH, the magenta bar shows the GaH measurement by \citet{Henley2013} for the southern galactic hemisphere.
                For ExG, the red bar shows the ExG measurement by \citet{Kolodzig2016} using the shallower survey \xbootes{}, and the blue line shows the expected contribution of unresolved point sources in our observation using the \LogNLogS\ of \citet{Luo2017}.
                These two reference values can be seen as upper and lower limits, respectively.
        }       
\end{center}            
\end{figure}                    
        
\Fref{CXB_spec_full} shows the energy spectrum of the CXB region with our best-fit model, described in \Sref{s:SpecFit}.
The posterior distributions of its free parameters are shown in \Fref{SpecFit_CXB}.
The best-fit values of its CXB-component parameters are listed in \Tref{t:CXBFit}
and their derived XSB in the $0.5-2.0\,\mathrm{keV}$ band are shown in \Fref{SpecFit_CXB_XSB}.
\Fref{ClA_spec_full} shows an energy spectrum of cluster A,
where our best-fit CXB model was used as fixed model.

\section{Profile projection onto the sky}  \label{a:proj}

We compute the projection along the LoS direction ($r_\parallel$) of the profile $Q(r_\mathrm{3D})$
via the abel transform $\mathcal{A}$:
\begin{align}
        \left(\mathcal{A}\;Q \right) (r_\perp) & =              
            \int\limits^\infty_{-\infty} \d r_\parallel \; Q(\sqrt{r_\parallel^2 + r_\perp^2}) \notag \\
            & \approx \int\limits^{r_\mathrm{max}}_{r_\perp} \d r_\mathrm{3D}  \; Q(r_\mathrm{3D}) \; \frac{2 \, r_\mathrm{3D} }{\sqrt{ r_\mathrm{3D}^2 - r_\perp^2 }} \label{eq:proj}
            \text{ ,}
\end{align}
with the coordinate convention:
$r_\mathrm{3D} = \sqrt{r_\parallel^2 + r_\perp^2}$ is the radial coordinate in 3D space ($r_\parallel$,$r_x$,$r_y$),
$r_\parallel$ is the coordinate in the LoS direction,
and $r_\perp = \sqrt{r_x^2 + r_y^2}$ is the radial coordinate in the projected plane ($r_x$,$r_y$).
The integration is done numerically via the Python package \textsc{PyAbel}\furl{https://github.com/PyAbel/PyAbel} \citep{PyAbel},
which also requires limiting the projection to a maximum radius ($r_\mathrm{max}$);
it is set to be three times the radius of the outermost profile bin ($r_\mathrm{max}=3\,r_\mathrm{last}$),
because for this case, our numerical solution is consistent within $\sim1.5\,\%$ with the analytical solution of the projection of $\epsilon_V(r_\mathrm{3D})$ for the isothermal beta-profile model with $\beta=2/3$.
As revealed later in our analysis, this is significant lower than our statistical uncertainty of measuring the XSB profile ($\approx20\,\%$) and the temperature profiles. 
We note that using much smaller values of $r_\mathrm{max}$, such as $r_\mathrm{max}=r_\mathrm{last}$, is not recommended, 
because it makes the numerical solution highly inaccurate for the outer part ($>r_\mathrm{last}/3$) of the profile.

\section{Results of the radial profile fit} \label{a:1DXfit}
\begin{table*} 
\begin{center}
\caption{\label{t:DevPara_r200} 
        Same as \Tref{t:DevPara} for $R_{200}$-related quantities.
        }
\begin{tabular}{cc|cc|cc|cc}
\hline
\multirow{2}{*}{Symbol} & \multirow{2}{*}{Unit} &  \multicolumn{2}{c}{Cluster A} & \multicolumn{2}{c}{Cluster B} & \multicolumn{2}{c}{Cluster C} \\
  &  & \multicolumn{1}{c}{Iso-$\beta$} & \multicolumn{1}{c|}{$\gamma$-NFW} & \multicolumn{1}{c}{Iso-$\beta$} & \multicolumn{1}{c|}{$\gamma$-NFW} & \multicolumn{1}{c}{Iso-$\beta$} & \multicolumn{1}{c}{$\gamma$-NFW}  \\
\hline
  $R_{200}$              & Mpc                           & $1.36\pm{0.05}$     & $1.35_{-0.05}^{+0.04}$ & $0.98_{-0.04}^{+0.05}$            & $0.99_{-0.04}^{+0.05}$ & $0.78_{-0.04}^{+0.05}$ & $0.79_{-0.04}^{+0.05}$ \\
 $\theta_{200}$         & arcmin                        & $4.3\pm{0.2}$       & $4.3\pm{0.1}$          & $3.11_{-0.12}^{+0.15}$            & $3.15_{-0.12}^{+0.16}$ & $2.5_{-0.1}^{+0.2}$    & $2.5\pm{0.1}$          \\
 $M_{t,200}$            & $10^{14}\,\mathrm{M_{\odot}}$ & $3.9_{-0.4}^{+0.5}$ & $3.9\pm{0.4}$          & \multicolumn{2}{c|}{$1.5\pm{0.2}$} & $0.74_{-0.11}^{+0.14}$ & $0.8\pm{0.1}$                                  \\
 $M_{\mathrm{gas},200}$ & $10^{13}\,\mathrm{M_{\odot}}$ & $4.0\pm{0.3}$       & $3.7\pm{0.3}$          & $2.5\pm{0.2}$                     & $2.6\pm{0.2}$          & $1.21_{-0.12}^{+0.16}$ & $1.27_{-0.13}^{+0.15}$ \\
\hline
\end{tabular}
\end{center}
\end{table*}

\begin{figure*}
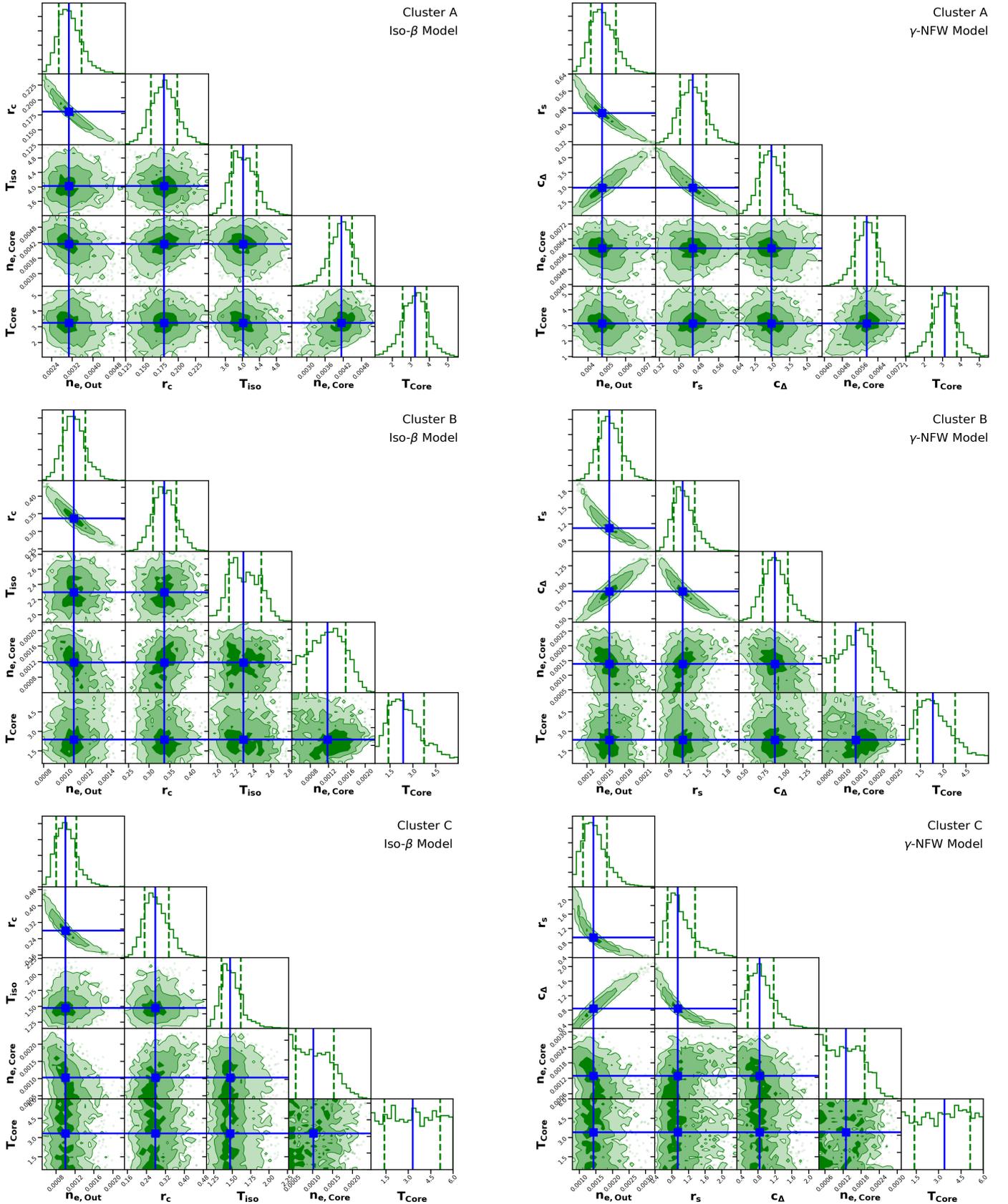

\begin{center} 

        \includegraphics[width=0.47\textwidth]{{\FigRootA _Corner_v3_Clu_A_M0}.png}
        \hfill
        \includegraphics[width=0.47\textwidth]{{\FigRootA _Corner_v3_Clu_A_M1}.png}
        
        \includegraphics[width=0.47\textwidth]{{\FigRootA _Corner_v3_Clu_B_M0}.png}
        \hfill
        \includegraphics[width=0.47\textwidth]{{\FigRootA _Corner_v3_Clu_B_M1}.png}
        
        \includegraphics[width=0.47\textwidth]{{\FigRootA _Corner_v3_Clu_C_M0}.png}
        \hfill
        \includegraphics[width=0.47\textwidth]{{\FigRootA _Corner_v3_Clu_C_M1}.png}
        
        \caption{\label{f:Corner_FitPara_1DFit_Clu} 
                Posterior distribution of free cluster-model parameters from the joint fit of radial $T$ and XSB profiles.
                Top, middle, and bottom rows correspond to clusters A, B, and C, respectively.
                \textit{Left column:} Isothermal beta-profile model (\Sref{s:Mod_IsoBeta}).
                \textit{Right column:} Polytropic NFW-profile model (\Sref{s:Mod_PolyNFW}).
                The blue solid lines show the median of each posterior distribution.
                For the 2D histograms, contours show the $1\sigma$, $2\sigma,$ and $3\sigma$ levels.
                For the 1D histograms, dashed lines show $1\sigma$ levels.
                The free background-model parameters are omitted for clarity but they do not show any degeneracy with the cluster-model parameters.
        }
\end{center}            
\end{figure*}

In \Fref{Corner_FitPara_1DFit_Clu} we show the posterior distribution of free cluster parameters of the best-fit models shown in \Fref{Fit_AllProf}.
We use those best-file models to derive $R_{200}$
and its associated angular scale ($\theta$), total hydrostatic mass ($M_t$), and gas mass ($M_\mathrm{gas}$), which are  
shown in \Tref{t:DevPara_r200}.
The same quantities for $R_{500}$ are shown in \Tref{t:DevPara}.

\end{appendix}

\end{document}